\documentclass[twocolumn,tighten,times]{aastex631} 
\usepackage{float}
\usepackage{graphicx}
\usepackage{amssymb}
\usepackage{amsbsy}
\usepackage{amsmath}
\usepackage{amsfonts}
\usepackage{mathrsfs}
\usepackage{color}
\usepackage{multirow}

\usepackage{bm}

\DeclareMathAlphabet{\mathbi}{OT1}{ptm}{bx}{it}
\SetMathAlphabet\mathbi{bold}{OT1}{ptm}{bx}{it}

\shorttitle{Mrk~1018}
\shortauthors{Lu et al.}

\begin{document}

\title{\bf\large A Short-lived Rejuvenation during the Decades-long Changing-look Transition in the Nucleus of Mrk 1018}

\author[0000-0002-2310-0982]{Kai-Xing Lu}
\affil{Yunnan Observatories, Chinese Academy of Sciences, Kunming 650011, China} 
\affil{Key Laboratory for the Structure and Evolution of Celestial Objects, Chinese Academy of Sciences, Kunming 650011, China} 

\author[0000-0001-5841-9179]{Yan-Rong Li}
\affil{Key Laboratory for Particle Astrophysics, Institute of High Energy Physics, Chinese Academy of Sciences, 19B Yuquan Road, Beijing 100049, China}

\author[0000-0003-4773-4987]{Qingwen Wu}
\affil{Department of Astronomy, School of Physics, Huazhong University of Science and Technology, Luoyu Road 1037, Wuhan, China}

\author[0000-0001-6947-5846]{Luis C. Ho}
\affil{Kavli Institute for Astronomy and Astrophysics, Peking University, Beijing 100871, China}
\affil{Department of Astronomy, School of Physics, Peking University, Beijing 100871, China}

\author[0000-0002-2419-6875]{Zhi-Xiang Zhang}
\affil{Department of Astronomy, Xiamen University, Xiamen, Fujian 361005, China}

\author[0000-0002-1530-2680]{Hai-Cheng Feng}
\affil{Yunnan Observatories, Chinese Academy of Sciences, Kunming 650011, China} 
\affil{Key Laboratory for the Structure and Evolution of Celestial Objects, Chinese Academy of Sciences, Kunming 650011, China} 

\author[0000-0003-3823-3419]{Sha-Sha Li}
\affil{Yunnan Observatories, Chinese Academy of Sciences, Kunming 650011, China} 
\affil{Key Laboratory for the Structure and Evolution of Celestial Objects, Chinese Academy of Sciences, Kunming 650011, China} 

\author[0000-0003-4280-7673]{Yong-Jie Chen}
\affil{Key Laboratory for Particle Astrophysics, Institute of High Energy Physics, Chinese Academy of Sciences, 19B Yuquan Road, Beijing 100049, China}

\author[0000-0002-0771-2153]{Mouyuan Sun}
\affil{Department of Astronomy, Xiamen University, Xiamen, Fujian 361005, China}

\author[0000-0002-7020-4290]{Xinwen Shu}
\affil{Department of Physics, Anhui Normal University, Wuhu, Anhui 241002, China}

\author[0000-0001-9457-0589]{Wei-Jian Guo}
\affil{National Astronomical Observatories, Chinese Academy of Sciences, 20A Datun Road, Chaoyang District, Beijing 100101, China}

\author[0000-0003-0202-0534]{Cheng Cheng}
\affil{Chinese Academy of Sciences South America Center for Astronomy, National Astronomical Observatories, CAS, Beijing 100101, China}

\author[0000-0003-4156-3793]{Jian-Guo Wang}
\affil{Yunnan Observatories, Chinese Academy of Sciences, Kunming 650011, China} 
\affil{Key Laboratory for the Structure and Evolution of Celestial Objects, Chinese Academy of Sciences, Kunming 650011, China} 

\author[0000-0001-7294-106X]{Dongchan Kim}
\affil{National Radio Astronomy Observatory, 520 Edgemont Road, Charlottesville, VA 22903, USA}

\author[0000-0001-9449-9268]{Jian-Min Wang}
\affil{Key Laboratory for Particle Astrophysics, Institute of High Energy Physics, Chinese Academy of Sciences, 19B Yuquan Road, Beijing 100049, China}

\author{Jin-Ming Bai}
\affil{Yunnan Observatories, Chinese Academy of Sciences, Kunming 650011, China}
\affil{Key Laboratory for the Structure and Evolution of Celestial Objects, Chinese Academy of Sciences, Kunming 650011, China}

\email{lukx@ynao.ac.cn}

\begin{abstract}
Changing-look active galactic nuclei (CL-AGNs), characterized by emerging or disappearing of 
broad lines accompanied with extreme continuum flux variability, have drawn much attention 
for their potential of revealing physical processes underlying AGN evolution. 
We perform seven-season spectroscopic monitoring on Mrk~1018, one of the earliest identified CL-AGN. 
Around 2020, we detect a full-cycle changing-look transition of Mrk~1018 within one year, 
associated with a nucleus outburst, which likely arise from the disk instability in the transition region 
between the outer standard rotation-dominated disk and inner advection-dominated accretion flow. 
Over the past forty-five years, the accretion rate of Mrk~1018 changed 1000 times and the maximum Eddington ratio reached 0.02. 
By investigating the relation between broad-line properties and Eddington ratio ($L_{\rm bol}/L_{\rm Edd}$), 
we find strong evidence that the full-cycle type transition is regulated by accretion. 
There exists a turnover point in the Balmer decrement, which is observed for the first time. 
The broad Balmer lines change from a single peak in Type 1.0-1.2 to double peaks in Type 1.5-1.8 
and the double-peak separation decreases with increasing accretion rate. 
We also find that the full width at half maximum (FWHM) of the broad Balmer lines obeys 
FWHM$\propto (L_{\rm bol}/L_{\rm Edd})^{-0.27}$, as expected for a virialized BLR. 
The velocity dispersion $\sigma_{\rm line}$ follows a similar trend in Type 1.5-1.8, 
but displays a sharp increases in Type 1.0-1.2, resulting in a dramatic drop of FWHM/$\sigma_{\rm line}$. 
These findings suggest that a virialized BLR together with accretion-dependent turbulent motions 
might be responsible for the diversity of BLR phenomena across AGN population. 
\end{abstract}


\section{Introduction} \label{intro}  
It is generally acknowledged that active galactic nuclei (AGNs) emit substantial energy from accretion of matter onto supermassive black holes (SMBHs) 
via an accretion disk (AD), leading to a wealth of  observational features across the electromagnetic spectrum (e.g., \citealt{Ho2008,Netzer2015}). 
The broad-line region (BLR), situated in the outer part of the accretion disk and the inner region of the dust torus, generates 
broad emission lines with varying strengths, velocity widths, and profile shapes, manifesting as a prominent feature in optical and ultraviolet spectra. The BLR plays a vital role in accurately determining SMBH masses and understanding AGN evolution (e.g., \citealt{Netzer2013,Elitzur2014,Zhou2019,Lu2022}). 
Nevertheless, the structure and origin of the BLR remain subjects of considerable debate.  

AGNs can be typically classified by the relative strength of the broad emission lines compared to narrow lines 
(e.g., the flux ratio of broad H$\beta$ to [O {\sc iii}]$\lambda$5007 lines), such as types 1.0, 1.2, 1.5, 
1.8, and 2.0 (\citealt{Osterbrock1977}), which respectively correspond to  H$\beta$/[O {\sc iii}]$>$5, 5$>$H$\beta$/[O {\sc iii}]$>$2, 
2$>$H$\beta$/[O {\sc iii}]$>$0.33,  H$\beta$/[O {\sc iii}]$<$0.33, and no broad emission lines (\citealt{Winkler1992}). 
The unification model posits that all AGNs share a similar structure, comprising an accreting SMBH surrounded by a dusty toroidal configuration (dust torus) and the diversity in AGN types arises from the different viewing angles (\citealt{Antonucci1993,Urry1995}). However, 
Changing-look events (CLEs) in AGNs, often referred as 
changing-look active galactic nuclei (CL-AGNs), exhibit significant continua variability along with transitions between AGN types over time scales much shorter than viscous time of accretion disks \citep{Khachikian1971,Penston1984,Cohen1986,Wang2024}, posing challenges to the viewing angle-dependent unification model.

CL-AGNs are further categorized into changing-obscuration AGNs (CO-AGNs) and changing-state AGNs (CS-AGNs) 
based on the distinct physical processes indicated by X-ray and optical/ultraviolet observations \citep{Ricci2023}. 
In CO-AGNs, CLEs observed in X-ray bands are primarily driven by changes in the column density along the line of sight \citep{Matt2003,Mereghetti2021}. 
In contrast, for CS-AGNs, CLEs in the optical and ultraviolet are typically linked to significant variations in the radiation field driven by accretion 
(e.g., \citealt{Sheng2017,Graham2020}). Recently, hundreds of CL-AGNs have been detected through the spectral characteristics of broad emission lines that either emerge (turn on) or vanish (turn off) (e.g., \citealt{MacLeod2016,Yang2018,Graham2020,Zeltyn2024,Guo2024a,Guo2024b}), 
yet only a limited number have shown recurring changing-look phenomena based on infrequent spectral observations over several decades (\citealt{Wang2024,Komossa2024}), such as Mrk~1018 \citep{Cohen1986,McElroy2016,Lyu2021}, Mrk~590 \citep{Denney2014}, NGC~4151 \citep{Shapovalova2010,Chen2023a,Feng2024}, NGC~2617 \citep{Moran1996,Shappee2014,Feng2021}, and NGC~1566 \citep{Oknyansky2019}. 
These changes correspond to transitions between different AGN types (from Type 1 to Type 1.2/1.5, Type 1.8/1.9, or Type 2 and vice versa). 
Significant efforts have been made to uncover the mechanisms behind these changes 
(e.g., \citealt{Husemann2016,Krumpe2017,Sheng2017,Kim2018,Sniegowska2020,Lyu2021,Liu2021,Veronese2024}). 
Notably, in the case of 1ES~1927+654, \cite{Li2022} suggested that the substantial change in accretion rate was 
due to a tidal disruption event (TDE; also see \citealt{Merloni2015}). 
However, the feeding mechanisms for SMBHs remain largely unresolved in most instances (e.g., \citealt{Gaspari2020}). 
Nonetheless, understanding of the changing-look process remains elusive, as a complete cycle of such transitions 
has not yet been thoroughly observed. 
Continuous monitoring of CL-AGNs could provide valuable insights into 
the AGN structure evolution, thereby unveiling the key physical mechanism of CLAGNs. 

Over the past few decades, numerous observational methods have been established to investigate the internal structure of AGNs. 
One notable method, spectroscopic reverberation mapping (SRM), has proven to be an effective approach for analyzing 
the geometry and dynamics of the BLR in AGNs, as well as for estimating the masses of SMBHs (e.g., \citealt{Blandford1982,Peterson1993}). 
SRM has revealed that the BLR exhibits a stratified structure, with the radii of low-ionization line emitters being approximately ten times larger than 
those of high-ionization lines \citep{Bentz2010}, aligning with the predictions of photoionization models. 
The Balmer decrement, which is the ratio of the fluxes of Balmer lines (H$\alpha$/H$\beta$), 
serves as an indicator of internal reddening in AGNs since the intrinsic Balmer decrement remains constant for a specific gas composition, 
such as the Case B recombination value of H$\alpha$/H$\beta$$\sim 2.74$ \citep{Osterbrock2006}. 
In observations, however, the typical Balmer decrement value tends to exceed the intrinsic value (e.g., around 3.1, \citealt{Lu2019b}). 
Additionally, the Balmer decrement can help examine the ionization effects within the BLR \citep{Korista2004,Wu2023,Li2024}. 
Monitoring variations in the continuum color is a widely employed technique for studying AGN variability and heating processes. 
Some studies have noted that the nucleus spectra of AGNs tend to be bluer when brighter from the multi-band 
variability data (e.g., \citealt{Cutri1985,Sakata 2011,Cai2016,Guo2016,Ren2022}), while mid-infrared spectra of CL-AGNs become redder when brighter \citep{Yang2018,Graham2020}. 
The large flux variations in CL-AGNs offer unique opportunity to probe the internal structures of AGNs by carefully studying the SRM, 
the Balmer decrement, and the color variations. 

To study the evolution of AGNs, particularly the accretion disk and broad-line region evolution, 
we initiated a long-term observational project utilizing the Lijiang 2.4m telescope to conduct spectroscopic monitoring 
of a carefully selected sample of well-known AGNs. This sample includes prominent reverberation mapping AGNs (e.g., NGC~5548, see \citealt{Lu2022}), changing-look AGNs (e.g., Mrk~1018, Mrk~590, Mrk~993, and Mrk~883), 
as well as candidates for SMBH binaries (e.g., SDSS J153636.22+044127.0, which is characterized by double-peaked broad emission lines). 
Mrk~1018, a merger galaxy classified as Hubble type S0 \citep{Koss2011,Veronese2024} at $z$=0.043, was one of the earliest confirmed CL-AGNs based on limited spectroscopic observations. It exhibited a transition from type~1.9 to 1.2 over approximately five years 
(from 1979 to 1986; \citealt{Cohen1986}), before reverting back to type 1.9 in 2015 \citep{McElroy2016}. 
Since 2017, we have conducted follow-up spectroscopic observations of this source. In this paper, 
we present findings of a full-cycle CLE that occurred around 2020, along with a full-cycle AGN type transition behavior observed in Mrk~1018. 

\begin{figure*}[htb]
\centering
\includegraphics[angle=0,width=0.98\textwidth]{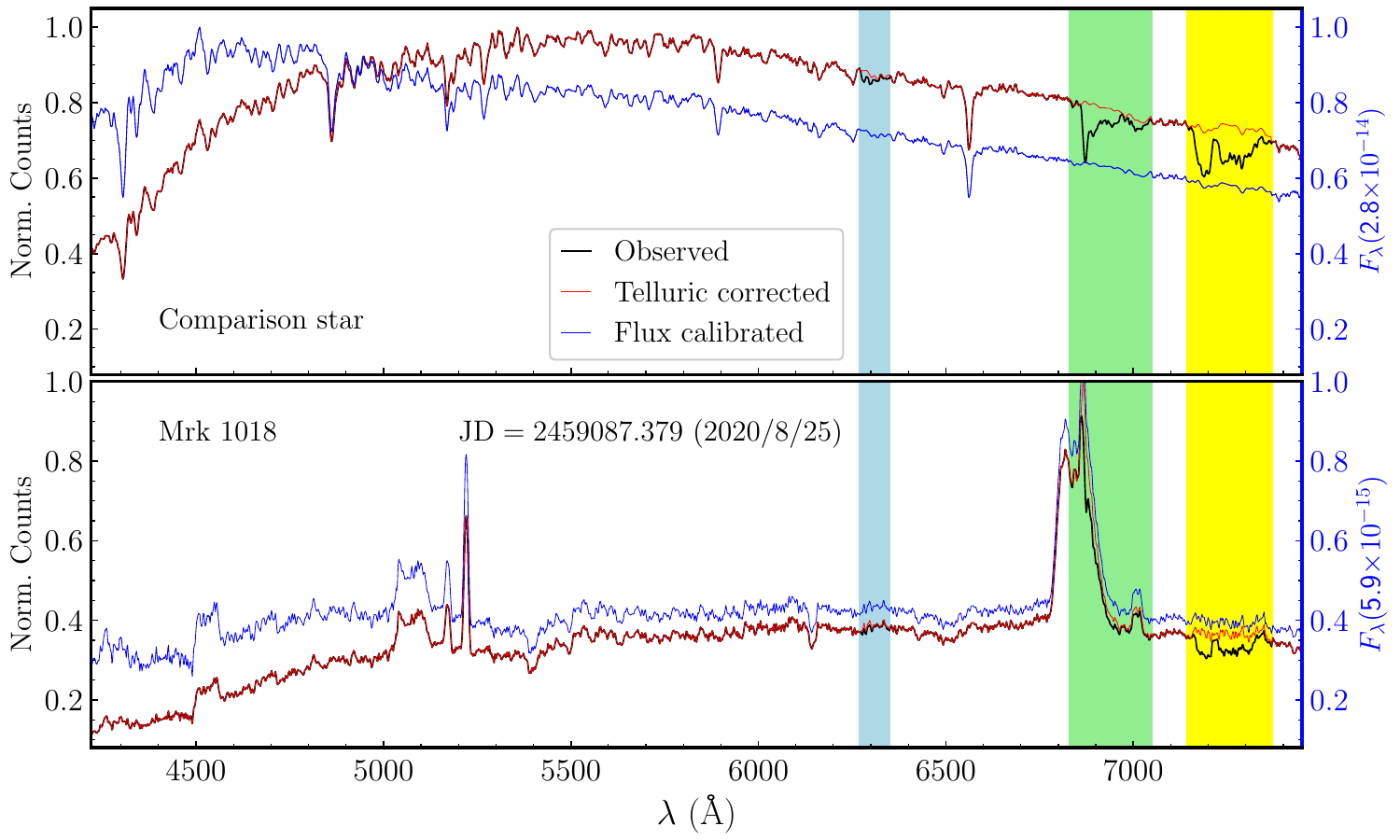}
\caption{
An example of the telluric correction and spectral flux calibration.
The top and bottom panels are for the comparison star and Mrk~1018, respectively.
In each panel, the black line shows the observed spectra (in counts, left vertical axis), the red line shows the telluric-corrected spectrum, and the blue line shows the flux-calibrated spectrum (in units of $\rm erg\,s^{-1}\,cm^{-2}\,\text{\AA}^{-1}$, right vertical axis). The three vertical colored bands represent telluric absorption windows. 
}
\label{speccali}
\end{figure*}

\begin{figure*}[htb]
\centering
\includegraphics[angle=0,width=0.98\textwidth]{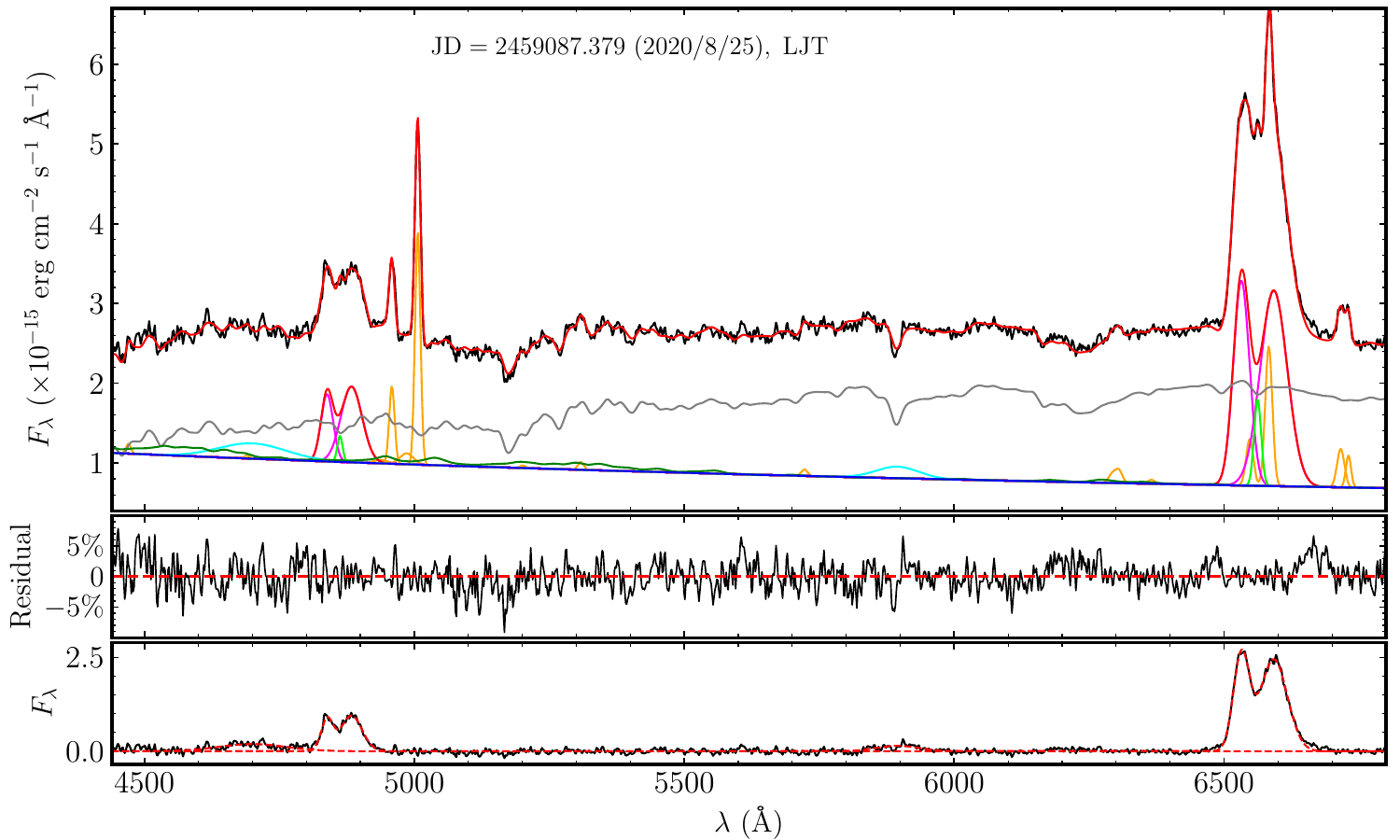}
\caption{
An example of multicomponent fitting and decomposition of the spectrum. 
The fitting model includes a power law ($f_{\lambda}\propto\lambda^{\alpha}$, $\alpha$ is the spectral index) for the AGN continuum (blue), 
an iron template for the iron multiplets (green), a stellar template for the host-galaxy starlight (gray), 
two double Gaussians for each broad Balmer line (${\rm H\beta}$ and ${\rm H\alpha}$, magenta), 
single Gaussian for each narrow Balmer line (${\rm H\beta}$ and ${\rm H\alpha}$, lime), 
single Gaussian for each Helium broad line (He~{\sc ii}~$\lambda$4686 and He~{\sc i}~$\lambda$5876, cyan), 
two double Gaussians for the [O~{\sc iii}] doublets (orange), 
single Gaussian for the [S~{\sc ii}] doublets and the [N~{\sc II}] doublets (orange), 
and other narrow emission lines. 
}
\label{specfit}
\end{figure*}

\section{Observation and Data} \label{obs}
\subsection{Optical Spectroscopy and Calibration} \label{specObs}
Long-term spectroscopic observations of Mrk~1018 were conducted utilizing the Yunnan Faint Object Spectrograph and Camera (YFOSC) 
attached to the Lijiang 2.4 m telescope (LJT). YFOSC is quipped with a back-illuminated 2048$\times$2048 pixel CCD, 
featuring a pixel size is 13.5 $\mu$m, a pixel scale is 0.283$''$ per pixel, a field-of-view is $10'\times10'$), 
and a series of filter and Grism \citep{Fan2015,Wang2019}. 
For spectroscopy of Mrk~1018, Grism 14 was employed, which spans wavelengths from approximately $\sim$3600~\AA\ to 7460~\AA\ 
and offers a dispersion of 1.8~\AA\,~pixel$^{-1}$. Following our earlier studies \citep{Lu2021a,Lu2021b,Lu2022}, 
a long slit with a projected width of  $2.5''$ was used, taking into account the average seeing conditions at the observatory. 

The spectroscopic monitoring of Mrk~1018 started on November 5, 2017, and to date, seven-season observations have been conducted 
from November 2017 to February 2024 (with Modified Julian Days ranging from 58063 to 60324). Similar to the long-term spectroscopic monitoring of NGC 5548 \citep{Lu2022}, we rotated the long slit within the field of view to simultaneously capture spectra of Mrk~1018 and a nearby stable comparison star. 
This observational approach is commonly employed in reverberation mapping campaigns, where the spectra of the comparison star 
can provide high-precision spectral calibration, including flux calibration and telluric absorption corrections (e.g., see \citealt{Hu2015,Lu2019a,Lu2021b} 
and below). Standard neon and helium lamps were used for wavelength calibration. 
In total, we acquired 96 spectroscopic observations over an approximate duration of seven years.

We processed the two-dimensional spectroscopic data using the standard procedures in {\tt PyRAF}. 
This involved bias subtraction, flat-field correction, wavelength calibration, and the removal of cosmic ray. 
Subsequently, all spectra were extracted with a consistent extraction window after subtracting the sky background. 
We opted for a relatively narrow extraction window of 20 pixels (5.7$^{\prime\prime}$), which aids in 
minimizing Poisson noise from sky background and increasing the signal-to-noise (S/N) ratio of the spectra. 
The sky background was estimated from two adjacent regions ($+7.4^{\prime\prime}\sim+14^{\prime\prime}$ 
and $-7.4^{\prime\prime}\sim-14^{\prime\prime}$) located on either side of the extraction window. 

The red side of ground-based optical spectrum is contaminated by the variable absorptions of the telluric atmosphere. 
In the observed spectrum of Mrk~1018, there are three telluric absorption windows, 
which are marked by vertical colored bands in Figure~\ref{speccali}. 
One of these windows (the green band) overlaps with the broad H$\alpha$ line.  
We first correct the telluric absorption and subsequently calibrate the spectral fluxes. 

Following the work of \cite{Lu2021b}, we processed the observed spectra of Mrk~1018 through a three-step calibration procedure. 
First, we determined the synthesis stellar template for the comparison star by aligning its observed spectrum 
with a library of synthesis stellar models (e.g., \citealt{Husser2013}). Next, 
we modeled the observed spectra of comparison star (in counts) using the chosen stellar template, 
while masking the telluric absorption bands during this modeling (i.e., the regions marked by color diagram in Figure~\ref{speccali}). 
We then derived the the telluric transmission spectra by dividing the observed spectrum of the comparison star by the modeled spectrum. 
This allowed us to use the telluric transmission spectra to adjust for the telluric absorption lines in each of the observed spectra. 
Figure~\ref{speccali} (left y-axis) displays the corrected spectra of telluric absorption (in red) alongside the observed spectra (in black), 
with the bottom panel representing the spectrum of Mrk~1018 and the top panel showing the spectrum of the comparison star. 
Finally, similar to the previous step, we generated a wavelength-dependent sensitivity function for each object/comparison star pair 
by comparing the telluric-corrected spectrum of the comparison star with the stellar template. 
This sensitivity function was then utilized to calibrate the spectrum of Mrk~1018 (see also \citealt{Lu2019a,Lu2021a,Lu2022}). 
The flux-calibrated spectra of Mrk~1018 (bottom panel) and its comparison star (top panel) are shown in blue in Figure~\ref{speccali} (right y-axis). 

Before our monitoring, there exist nine spectra concurrently covering the broad H$\beta$ and H$\alpha$ lines (\citealt{Kim2018,Osterbrock1981,Cohen1986,Abazajian2009,Jones2009,Trippe2010,McElroy2016}), 
which enable us to investigate the extensive long-term spectral evolution  of Mrk 1018 (see Section~\ref{blrph}). 
We obtained these early spectra from \cite{Kim2018} 
and recalibrated them by assuming the same [O~{\sc iii}]$\lambda5007$ flux as our spectra. 
The first spectra of Mrk~1018 was taken in the 1970s, meaning that our spectroscopic baseline spans more than 45 years. 

\begin{figure*}[htb]
\centering
\includegraphics[angle=0,width=0.7\textwidth]{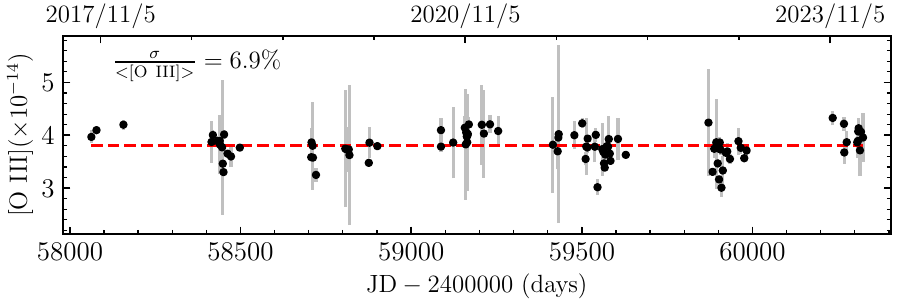}
\caption{
The light curve of the [O~{\sc iii}]~$\lambda$5007 in the LJT observation period.
The red dashed line represents the mean flux (3.8$\times10^{-14}~{\rm erg\,s^{-1}\,cm^{-2}}$). 
}
\label{oiii}
\end{figure*}

\begin{figure*}[t]
\centering
\includegraphics[angle=0,width=0.9\textwidth]{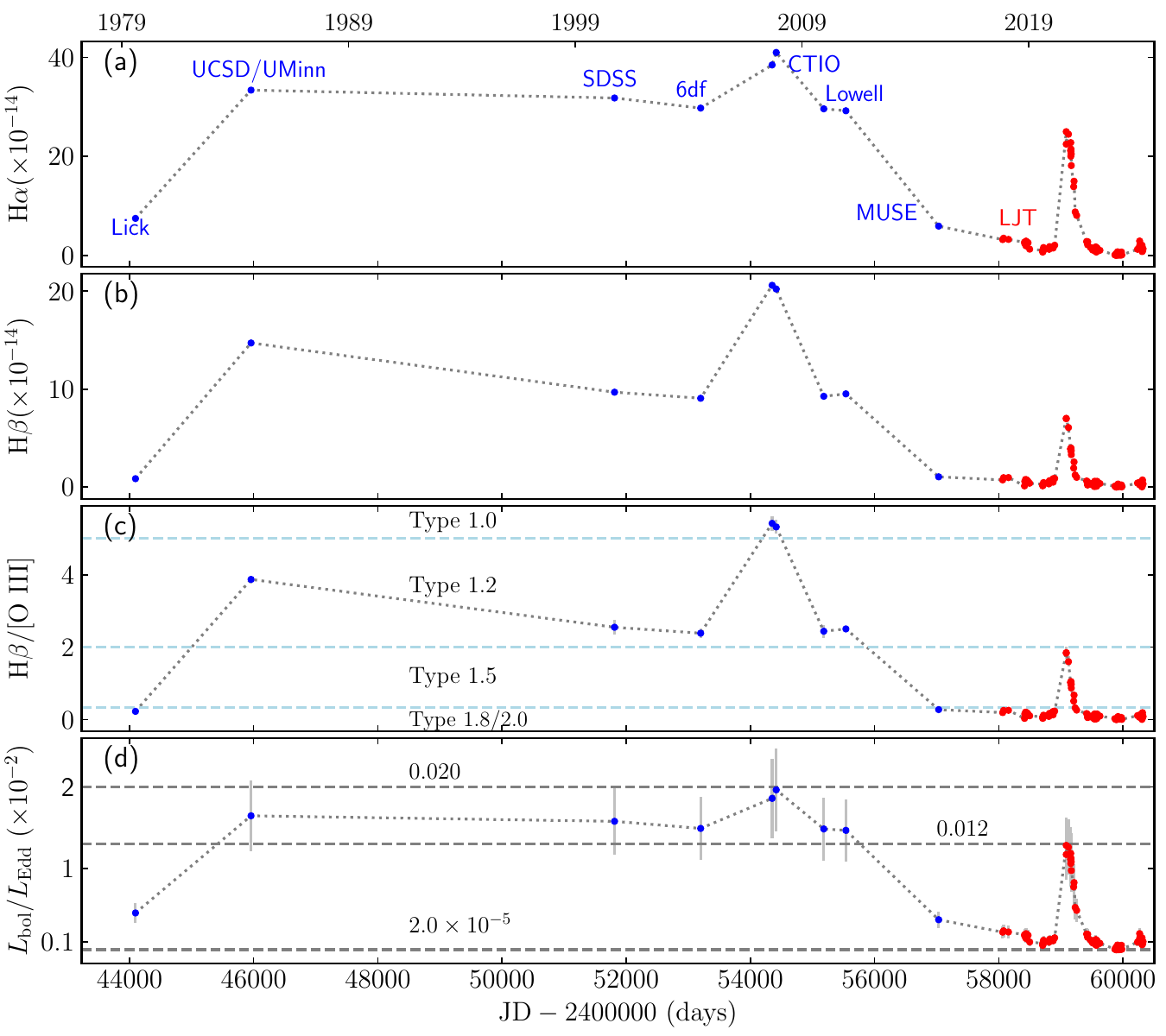}
\caption{
Long-term variability of Mrk~1018 from 1979 to 2024 for (a) the H$\alpha$ flux, (b) the H$\beta$ flux, (c) the flux ratio H$\beta$/[O~{\sc iii}]$\lambda$5007 (indicating the type transition), and (d) the Eddington ratio. 
The H$\alpha$ and H$\beta$ fluxes are in units of $\rm erg\,s^{-1}\,cm^{-2}$. The earlier data are in blue and the LJT data are in red.
}
\label{longvari}
\end{figure*}

\begin{figure}[htb]
\centering
\includegraphics[angle=0,width=0.48\textwidth]{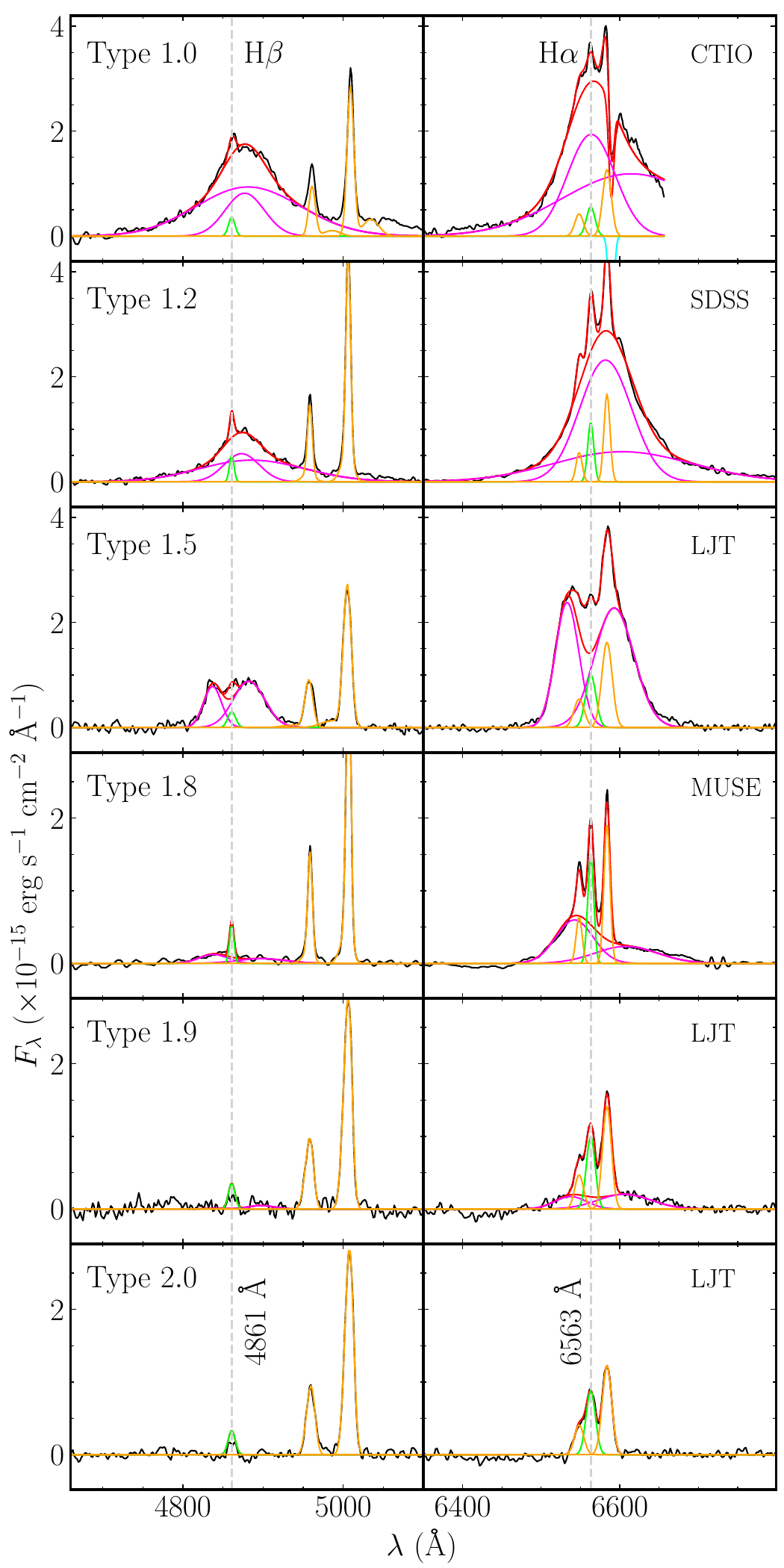}
\caption{\footnotesize
Evolution of the H$\beta$ and H$\alpha$ line profiles. 
Each row of panels displays the characteristic profiles of different types, with the broad H$\beta$ in the left and H$\alpha$ in the right.
The black curve shows the original data, the magenta curves show the broad-line components, 
the orange curves show [O~{\sc iii}]~$\lambda4959/\lambda5007$ lines in the H$\beta$ region 
and [N~{\sc II}]~$\lambda6548/\lambda6583$ lines in the H$\alpha$ region, and
the lime curves show the narrow H$\beta$ and H$\alpha$. We add an absorption component in the fitting of the CTIO spectra (top right panel, in cyan).  
The types, including 1.0, 1.2, 1.5, 1.8, 1.9 and 2.0 (no broad H$\beta$ and H$\alpha$ lines), 
are marked in the left panels and the spectral sources are marked in the right panels. 
The wavelength shifts for the broad H$\beta$ and H$\alpha$ regions are corrected using the narrow H$\beta$ and H$\alpha$ lines, respectively.
}
\label{belevo}
\end{figure}

\subsection{Spectral Fitting and Measurements} \label{specAnalysis}
The spectral fitting scheme is commonly adopted in spectral analysis of AGNs to separate the blending components 
(e.g., \citealt{Dong2011,Park2012,Guo2014,Barth2015}). 
In line with previous fitting routine (e.g., \citealt{Lu2022}), we conducted the spectral fitting and multicomponent decomposition 
to extract the spectral characteristics of Mrk 1018 (see Figure~\ref{specfit}), including variability and velocity shifts of broad emission lines. 
To cover the broad H$\beta$ and H$\alpha$ regions while minimizing degrees of freedom, 
we set the fitting window from 4440~\AA\ to 6800~\AA\ in rest frame for all Mrk~1018 spectra. 
The fitting includes the following components: 
(1) A power law ($f_{\lambda}\propto\lambda^{\alpha}$, $\alpha$ is the spectral index) representing the AGN continuum.
(2) An iron template from \cite{Boroson1992} for the iron multiplets. 
(3) A stellar template with an age of 11 Gyr and metallicity of Z = 0.05 from \cite{Bruzual2003} to account for the host-galaxy starlight. 
(4) Double Gaussians for each broad Balmer lines (H$\beta$ and H$\alpha$). 
In practice, we initially attempted to fit the broad Balmer line using double Lorentzians 
(Lorentzian was employed for the broad Balmer line of Narrow-Line Seyfert 1 galaxies, see \citealt{Veron2001}), 
but found the fits are inferior compared to those using double Gaussians for the double-peak broad lines. 
Therefore, we chose to use double Gaussians to model the broad Balmer lines. 
(5) Two sets of double Gaussians for the [O~{\sc iii}] doublets $\lambda5007/\lambda$4959. 
(6) A single Gaussian for each narrow emission line. 
(7) Four sets of single Gaussian for the [S~{\sc ii}] doublets and the [N~{\sc ii}] doublets. 
All components were fitted simultaneously across the fitting window using the MPFIT package \citep{Markwardt2009}, 
which employs the Levenberg$-$Marquardt algorithm for $\chi^{2}$-minimization. During the spectral fitting for each spectrum, 
the flux ratio of the [O~{\sc iii}] doublets was fixed to the theoretical value of 3 (e.g., \citealt{McGill2008,Hu2015}); 
all the narrow emission lines were tied to the same velocity and shift, 
while the rest of the fitting parameters were allowed to vary. Various host-galaxy templates from \cite{Bruzual2003} were also evaluated and 
the template with an age of 11 Gyr and a metallicity Z = 0.05 provided a reasonable fit for the spectral index of AGNs continuum ($\sim\lambda^{-1.5}$) 
and the stellar absorption lines. While several iron templates have been proposed for 
fitting the optical or ultraviolet spectra of AGNs (e.g., \citealt{Boroson1992,Veron2004,Kova2010,Park2022}), 
we found that the spectral fitting of Mrk~1018 is not sensitive to the choice of the specific iron template. 

Next, we focus on the measurements of spectral characteristics. 
The broad H$\beta$ and H$\alpha$ lines fluxes for each spectrum are measured from the optimal fitted models 
and are tabulated in Table~\ref{tab0} along with the uncertainties (including Poisson errors and systematic errors). 
The fluxes of the AGNs continuum can be measured from the best-fitted power-law component at 5100~\AA. 
During our observation period, the pseudo-continuum is dominated by the starlight from the host galaxy.The mild degeneracy between power-law continuum and host-galaxy component will cause scatter to the obtained
AGN continuum at 5100~\AA. Therefore, we just use the averaged AGN continuum flux at 5100~\AA~as a a reference for 
the intercalibration of different photometric light curves (see Section~\ref{photodata}). 
We measure the flux of [O~{\sc iii}]$\lambda5007$ line from the best-fitted model and report the result in Figure~\ref{oiii}, which exhibits a scatter of 6.9\%. 
This level of scatter indicates a sufficient accuracy in our spectral calibration. 

\begin{deluxetable}{lcc}
\tablecolumns{3}
\tabletypesize{\footnotesize}
\setlength{\tabcolsep}{4pt}
\tablewidth{4pt}
\tablecaption{The fluxes of broad Balmer lines \label{tab0}}
\tablehead{
\colhead{JD}                                    &
\colhead{H$\beta$}                         &
\colhead{H$\alpha$}                      \\
\colhead{(-2,400,000~days)} &
\colhead{($10^{-14}~{\rm erg\,s^{-1}\,cm^{-2}}$)} &
\colhead{($10^{-14}~{\rm erg\,s^{-1}\,cm^{-2}}$)} 
}
\startdata
\multicolumn{3}{c}{Previous observations}\\ \cline{1-3}
44097.500 &$ 0.855 \pm 0.070 $&$ 7.453 \pm 0.108 $\\
45958.500 &$ 14.711 \pm 0.124 $&$ 33.368 \pm 0.185 $\\
51812.500 &$ 9.685 \pm 0.139 $&$ 31.777 \pm 0.138 $\\
53199.500 &$ 9.063 \pm 0.185 $&$ 29.738 \pm 0.153 $\\
54350.500 &$ 20.599 \pm 0.146 $&$ 38.452 \pm 0.177 $\\
54416.500 &$ 20.221 \pm 0.142 $&$ 40.974 \pm 0.157 $\\
55182.500 &$ 9.268 \pm 0.079 $&$ 29.589 \pm 0.107 $\\
55538.500 &$ 9.514 \pm 0.083 $&$ 29.190 \pm 0.118 $\\
57032.500 &$ 1.041 \pm 0.059 $&$ 5.889 \pm 0.085 $\\
\hline
\multicolumn{3}{c}{Our seven-season monitoring}\\ \cline{1-3}
58063.136 &$ 0.725 \pm 0.114 $&$ 3.174 \pm 0.284 $\\
58078.118 &$ 0.956 \pm 0.103 $&$ 3.448 \pm 0.282 $\\
58157.048 &$ 0.966 \pm 0.109 $&$ 3.208 \pm 0.284 $\\
58415.239 &$ 0.116 \pm 0.121 $&$ 2.635 \pm 0.286 $
\enddata
\tablecomments{\footnotesize
The spectral sources and measurements refer to Section~\ref{specObs} and \ref{specAnalysis}, respectively. 
(This table is available in its entirety in machine-readable form.)
}
\end{deluxetable}

In spectral fitting described above, the broad H$\beta$ and H$\alpha$ lines in each observation are well modeled by two Gaussian components, 
This allows us to extract the characteristics of the broad Balmer lines from the optimal models. 
After correcting the wavelength shift of broad Balmer lines, which usually arise from varying seeing and mis-centering, 
using its narrow lines as wavelength reference, we measure the peak separations of two Gaussian components 
(including broad H$\beta$ and H$\alpha$ lines) based on the differences in peak wavelengths. 
Additionally, we measure the line widths including full-width at half-maximum (FWHM) and 
velocity dispersion ($\sigma_{\rm line}$) for broad H$\beta$ and H$\alpha$ lines. 
To reasonably determine the FWHM of the double-peaked emission line, we identify a blue-side peak and a red-side peak within the line profile, 
and the FWHM is calculate from the maximum wavelength separation of these peaks at each half-maximum, as described by \cite{Peterson2004}. 
In practice, instrument broadening is coupled with varying atmospheric (seeing) broadening, 
so we estimate the total broadening of the broad emission line for each observation 
by comparing the width of the [O~{\sc iii}]~$\lambda$5007 line with that from the SDSS spectrum, 
which is then used to derive the broadening-corrected line width. 
These measurements will be used to investigate the broad-line properties in subsequent analyses. 

\subsection{Optical Photometry and Intercalibration} \label{photodata}
To obtain the long-term variability of Mrk 1018, we compiled the photometry data from the public archival database of Zwicky Transient Facility (ZTF; \citealt{Graham2019}) and Asteroid Terrestrial-impact Last Alert System (ATLAS; \citealt{Tonry2018}). We use the ZTF $g$- and $r$-band data, along with the ATLAS $c$- and $o$-band data, and combine them using the Python package PyCALI\footnote{\url{https://github.com/LiyrAstroph/
PyCALI}} (\citealt{LYR2014}), which employs a multiplicative factor and an additive factor to intercalibrate different sources of light curves. We adopt ZTF-$g$ as the reference dataset and align other datasets with it.

\subsection{Mid-infrared Photometry} \label{MIphotodata} 
Mrk 1018 is also covered by the Wide-field Infrared Survey Explorer (WISE; {\citealt{Wright2010}) survey. 
We retrieved the photometric data with 
the following flags: the best frame image quality score ($qi_{-}fact=1$), no contamination of scattered light from the moon ($moon_{-}masked=0$), 
the larger South Atlantic Anomaly separation ($saa_{-}sep>=5$), and spurious detections excluded ($cc_{-}flags=0$). 
When necessary, we convert $W1$ and $W2$ magnitudes into flux densities by adopting zero-point flux densities 
$F_{\lambda}(W1=0)=8.03\times10^{-12}~{\rm erg~s^{-1}cm^{-2}\AA^{-1}}$ and
$F_{\lambda}(W2=0)=2.43\times10^{-12}~{\rm erg~s^{-1}cm^{-2}\AA^{-1}}$ \citep{Jarrett2013}. 
The resulting fluxes and magnitudes are rebinned every 6 months. 

\begin{deluxetable*}{lccccccc}
\tablecolumns{4}
\tabletypesize{\footnotesize}
\setlength{\tabcolsep}{4pt}
\tablewidth{4pt}
\tablecaption{The broad H$\alpha$ line fluxes and derived parameters \label{tab1}}
\tablehead{
\colhead{}  &
\multicolumn{3}{c}{1979-2024}         &
\colhead{}  &
\multicolumn{3}{c}{Outburst around 2020}    \\ 
\colhead{}  &
\multicolumn{3}{c}{\bf (JD 2,444,097 to 2,460,324)}         &
\colhead{}  &
\multicolumn{3}{c}{\bf (JD $\sim$2,458,900 to $\sim$2,459,300)}    \\ \cline{2-4} \cline{6-8} 
\colhead{Parameters}                           &
\colhead{Minimum}                                     &
\colhead{H$\beta$/[O~{\sc iii}]$\lambda$5007=0.1}                                     &
\colhead{Maximum}   &
\colhead{}  &
\colhead{Before the outburst}                     &
\colhead{Peak}                                     &
\colhead{After the outburst}                        \\
\colhead{(1)}  &
\colhead{(2)}  &
\colhead{(3)}  &
\colhead{(4)}  &
\colhead{}  &
\colhead{(5)}  &
\colhead{(6)}  &
\colhead{(7)}  
}
\startdata
H$\alpha$ Flux (10$^{-14}$ erg s$^{-1}$ cm$^{-2}$)  &0.01 &0.75&40.97  && 1.32   & 24.97     & 1.41  \\ 
H$\alpha$ Luminosity (10$^{40}$ erg s$^{-1}$)         &0.05 &3.26&177.84 && 5.73    & 108.39  & 6.12   \\ 
Luminosity at 5100~\AA~(10$^{42}$ erg s$^{-1}$)    &0.004 &1.26&39.21 && 2.04    & 25.70    & 2.16  \\ 
Bolometric Luminosity (10$^{43}$ erg s$^{-1}$)         &0.04 &1.23&38.43  && 2.00    & 25.08    & 2.12   \\ 
Eddington ratio                                                           &2.0$\times10^{-5}$ &6.1$\times10^{-4}$ & 0.020 && 1.0$\times10^{-3}$ & 0.012    & 1.0$\times10^{-3}$    
\enddata
\tablecomments{\footnotesize
Columns (2-4) are the relevant parameters observed between 1979 and 2024 (JD 2,444,097 to 2,460,324), while columns (5-7) represent the outburst phase around 2020 (JD $\sim$2,458,900 to $\sim$2,459,300, also see Table~\ref{tab2}). 
The optical luminosity of $L_{\rm 5100}$=$2.39\times10^{43}(L_{\rm H\alpha}/10^{42})^{0.86} ~\rm erg~s^{-1}$ was given by \cite{Greene2005}, 
where $L_{\rm H\alpha}$ denotes the luminosity of broad H$\alpha$ line. 
Using the relation of $L_{\rm bol}$=$9.8L_{\rm 5100}$ \citep{McLure2004}, we derive the bolometric luminosity of $L_{\rm bol}$=$2.34\times10^{44}(L_{\rm H\alpha}/10^{42})^{0.86}$~erg~s$^{-1}$. 
Here, the cosmology with $H_0=72{\rm~km~s^{-1}~Mpc^{-1}}$, $\Omega_{\Lambda}=0.7$, 
and $\Omega_{\rm M}=0.3$ is adopted. 
}
\end{deluxetable*}

\section{Data analysis and results} \label{dataanalysis}
The light curves for broad H$\alpha$ and H$\beta$ lines, along with the spectral sources (panels~a and b), 
and the variations of AGN type characterized by line ratio of broad H$\beta$ and [O~{\sc iii}]$\lambda$5007 (panel~c) are illustrated in Figure~\ref{longvari}. 
By applying the relationship between the optical luminosity ($L_{\rm 5100}$) and broad H$\alpha$ luminosity ($L_{\rm H\alpha}$) 
established by  \cite{Greene2005}, represented as $L_{\rm 5100}$=$2.39\times10^{43}(L_{\rm H\alpha}/10^{42})^{0.86}$~erg~s$^{-1}$, 
we can obtain the optical luminosity for each observation. 
Furthermore, using the relation $L_{\rm bol}$=$9.8~L_{\rm 5100}$ \citep{McLure2004}, we derive the bolometric luminosity of 
$L_{\rm bol}$=$2.34\times10^{44}(L_{\rm H\alpha}/10^{42})^{0.86}$~erg~s$^{-1}$, 
which allows us to estimate the bolometric luminosity for each observation. 
Finally, using the Eddington luminosity $L_{\rm Edd}$=$1.26\times10^{38}(M_{\rm BH}/M_{\sun})$~erg~s$^{-1}$, 
we estimate the Eddington ratio ($L_{\rm bol}/L_{\rm Edd}$) from the derived bolometric luminosity, and show its variations in panel (d) of Figure~\ref{longvari}. 
The details on the estimation of SMBH mass and Eddington luminosity refer to Section~\ref{sizeBHmass}. 
To inspect the change of broad emission line, we select and present the typical emission line spectra in Figure~\ref{belevo}, 
focusing on the broad H$\alpha$ and H$\beta$ regions and covering types 1.0, 1.2, 1.5, 1.8, 1.9 and 2.0. 
In columns (2-4) of Table~\ref{tab1}, we list the typical values for the broad H$\alpha$ line fluxes and related parameters, 
including the minimum and maximum values, as well as the values as H$\beta$/[O~{\sc iii}]$\lambda$5007=0.1. 
This ratio of 0.1 nearly represents the detection threshold for broad H$\beta$ lines 
due to their flux being comparable to that of narrow H$\beta$ lines. 
Additionally, we calculate the logarithm ratios of narrow line ratios using all observations from LJT, specifically [O~{\sc iii}]$\lambda5007$/H$\beta$ 
and [N~{\sc II}]$\lambda6584$/H$\alpha$, for each observation, yielding $0.18\pm0.03$ and $1.01\pm0.01$, respectively, 
indicating Mrk~1018 is a Seyfert galaxy (\citealt{Kewley2001}). 
These findings reveal that the Eddington ratio of Mrk~1018 has varied by a factor of 1000 over the last 45 years of observation, 
implying a significant modification of the central engine, and reflecting a transition from passive to active states in the SMBH. 
Consequently, Mrk~1018 serves as an exemplary target for investigating the physical processes underlying AGN evolution. 

\begin{figure*}[htb]
\centering
\includegraphics[angle=0,width=0.49\textwidth]{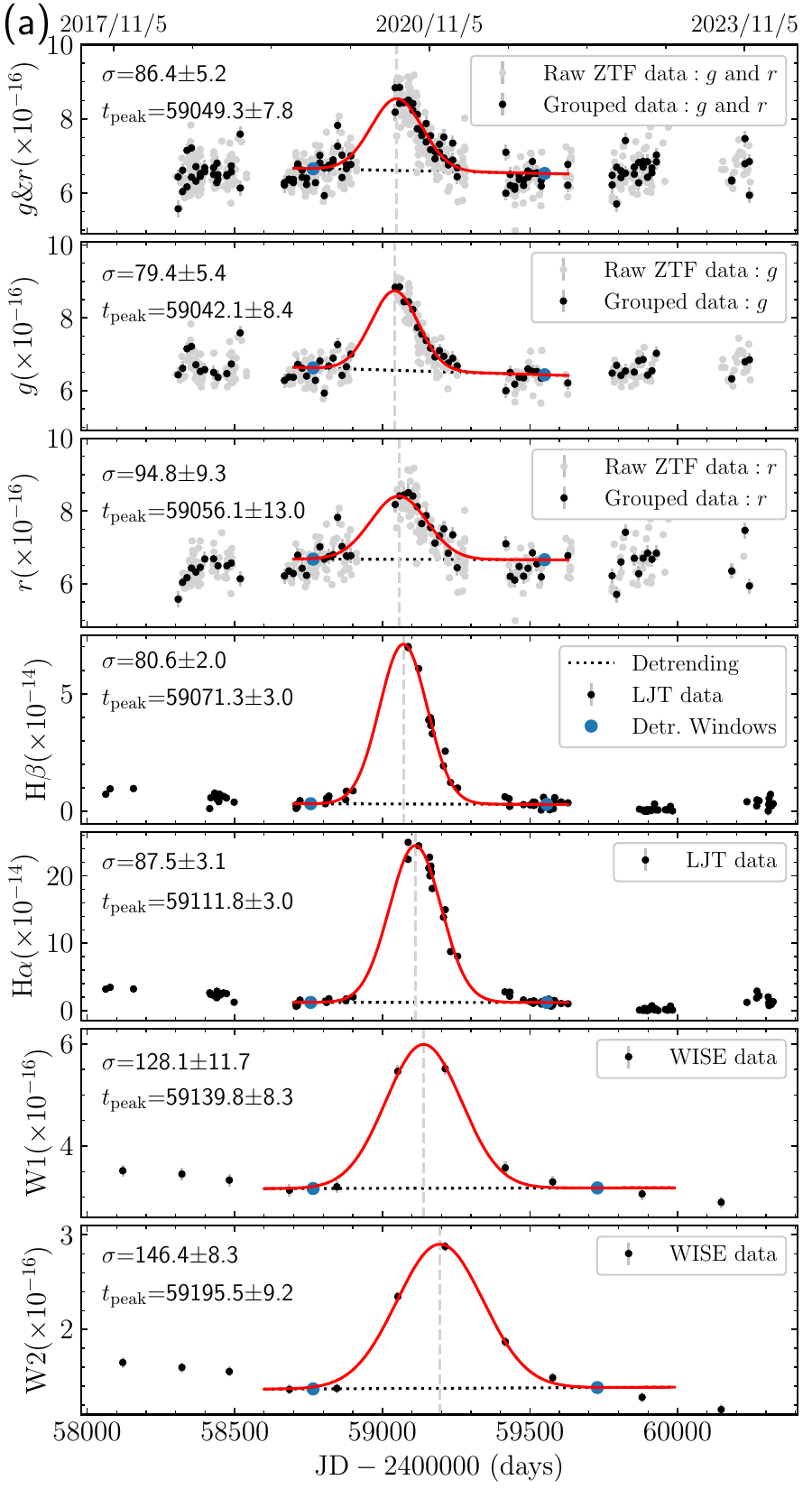}
\includegraphics[angle=0,width=0.49\textwidth]{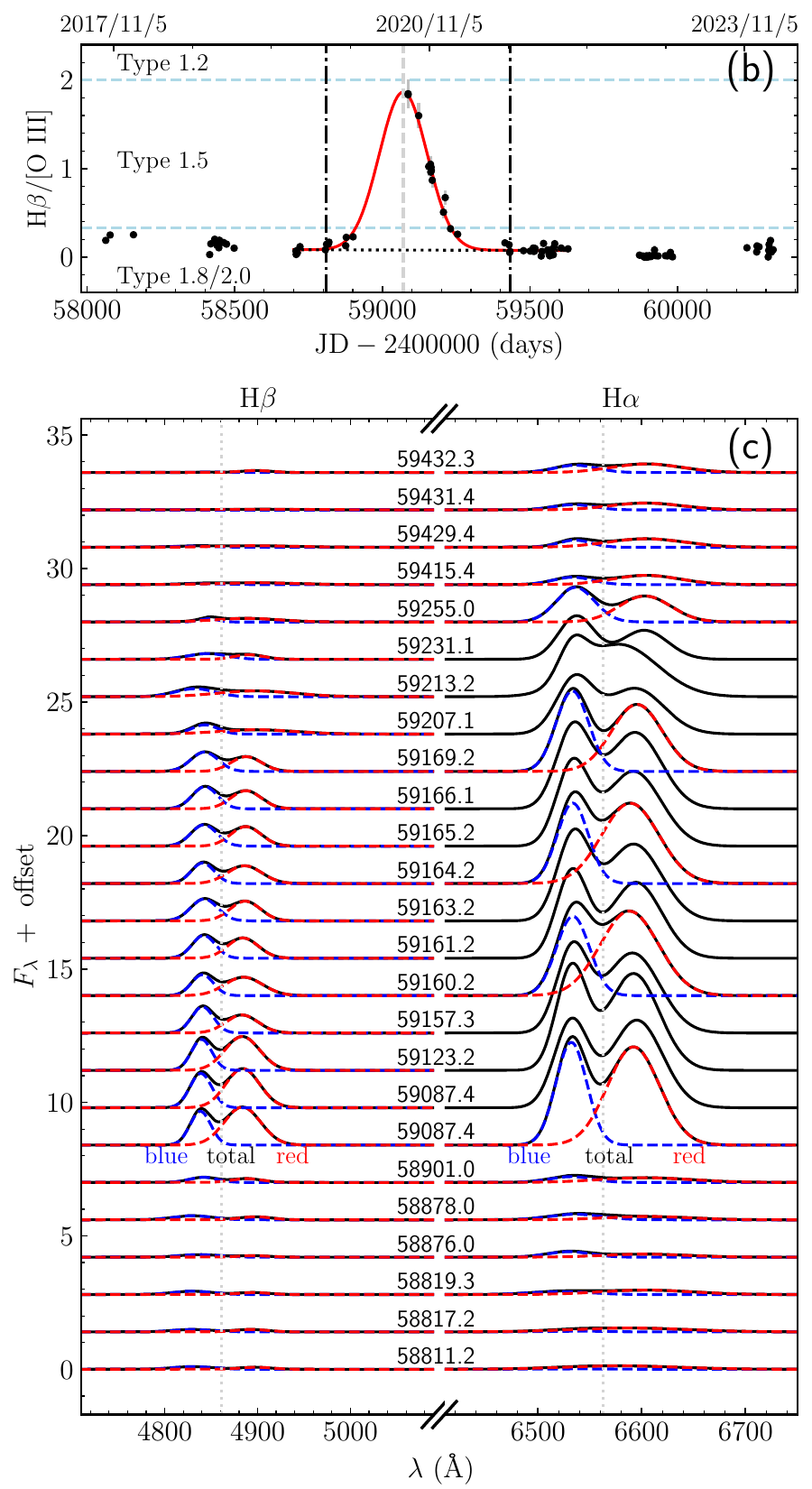}
\caption{
Multi-scale outbursts accompanied by a short-lived yet full-cycle changing-look transition in Mrk~1018. 
Panel~(a) displays the light curves of ZTF photometry (combined $g\&r$-, $g$-, and $r$-band, in units of $\rm erg\,s^{-1}\,cm^{-2}\,\text{\AA}^{-1}$), 
the broad Balmer lines (H$\beta$ and H$\alpha$, in units of $\rm erg\,s\,cm^{-2}$), as well as the WISE photometry ($W1$- and $W2$-band, in units of $\rm erg\,s^{-1}\,cm^{-2}\,\text{\AA}^{-1}$). 
After subtracting a linear trend determined from the adjacent regions around the outburst (blue circles indicating the centers of chosen detrending windows, 
and dotted lines representing linear trends), the light-curve profiles of the outburst are fitted using a Gaussian (in red), 
yielding a typical $\chi^2/{\rm dof}\sim 2$. The vertical dashed line shows the peak position of the outburst ($t_{\rm peak}$). 
The fitting parameters, including $\sigma$ (days) and $t_{\rm peak}$ (-2,400,000~days), are noted in each sub-panel. 
Panel~(b) presents the variations in AGN type, where the type-transition profile is also modeled by a Gaussian (in red) that 
shares the same parameters as the broad H$\beta$ modeling. 
The broad H$\beta$ and H$\alpha$ line profiles within vertical dot-dashed lines are showed in panel (c). 
Panel~(c) shows the fitted broad H$\beta$ and H$\alpha$ line profiles for the outburst period, 
they are marked by the observation time of JD (-2,400,000~days), which increases from the bottom to the top. 
The pronounced double-peaked shape highlights the broad-line feature, with the total broad lines (in black) representing the best fitted model, 
composed of a blue-shifted Gaussian (in blue dashed lines) and a red-shifted Gaussian (in red dashed lines). 
}
\label{cle1}
\end{figure*}

\begin{figure*}[htb]
\centering
\includegraphics[angle=0,width=0.47\textwidth]{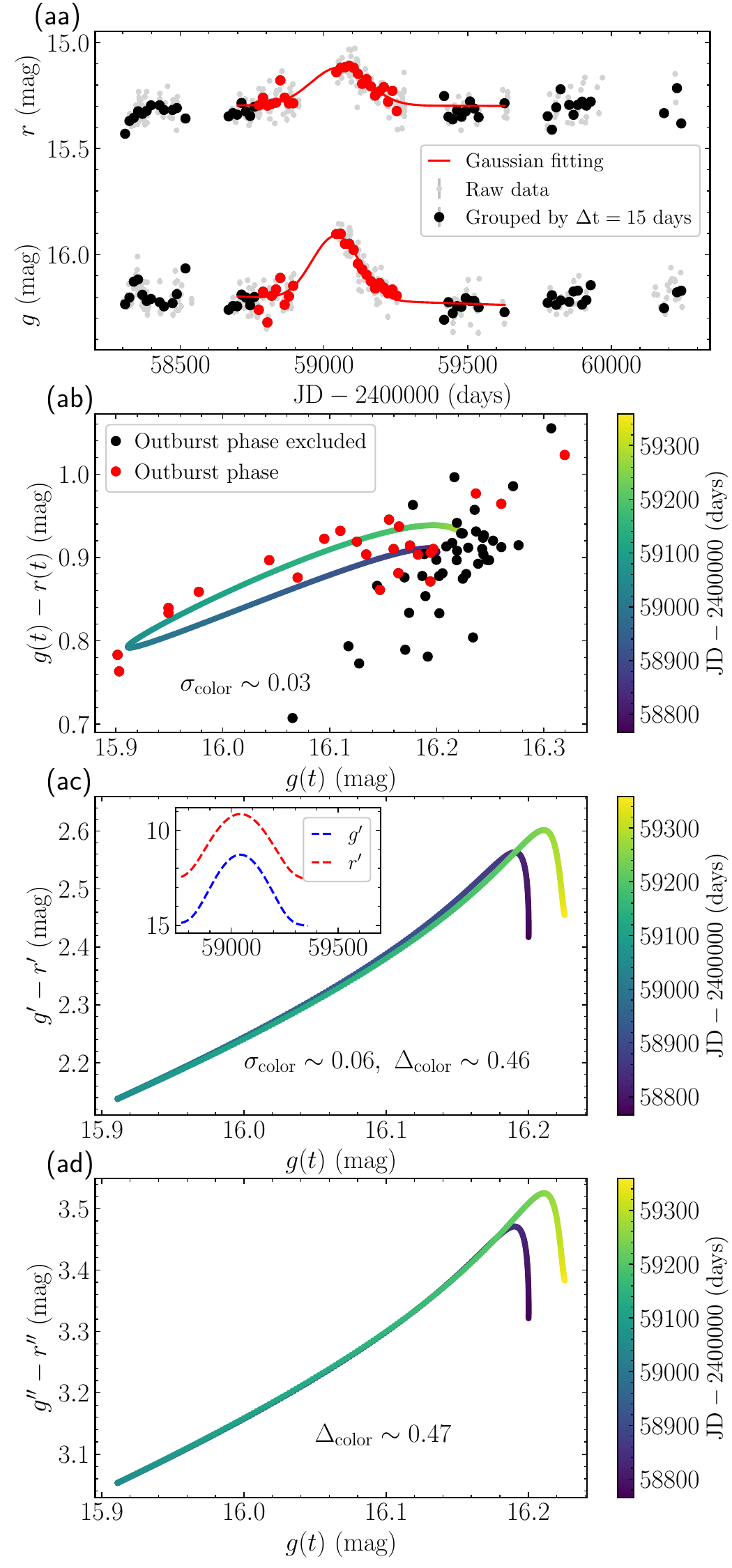}
\includegraphics[angle=0,width=0.47\textwidth]{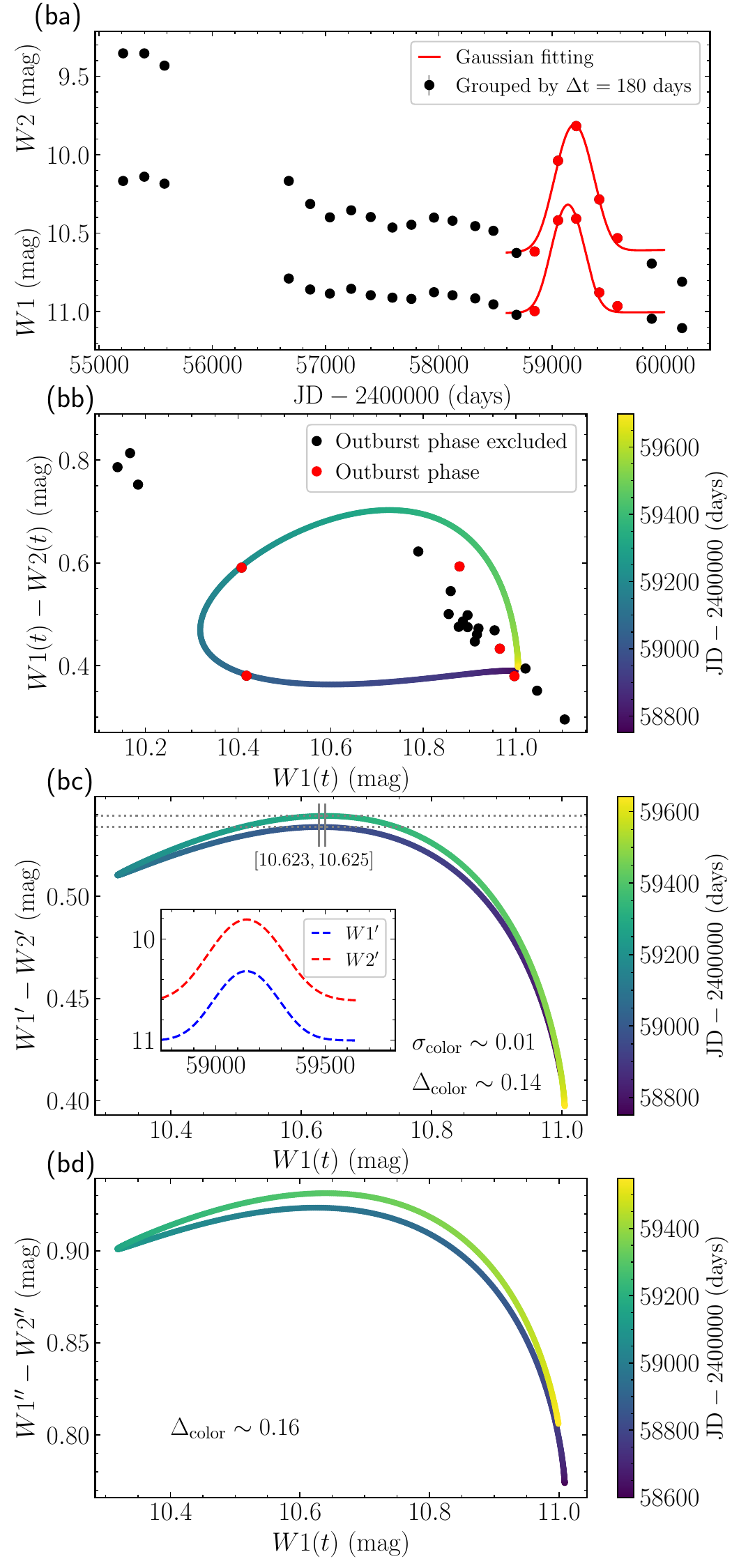}
\caption{
Optical and mid-infrared color$-$magnitude relations based on the Gaussian fitting model (see Appendix~\ref{color}). 
The left panels~(aa-ad) are for the optical bands, while the right panels~(ba-bd) are for the mid-infrared bands. 
The first row (panels aa and ba) displays the observed (points) alongside the modeled light curves (red lines), with the outburst phase highlighted in red. 
We calculate the optical and mid-infrared color$-$magnitude relations in three cases and present in the next three rows, respectively. 
In the second row (panels ab and bb), 
the color$-$magnitude relations are generated directly using the grouped data (where the outburst phase is highlighted in red points) 
and fitting model of the outburst phase (in colored lines). 
In the third row (panels ac and bc), 
because the ZTF $g$-band covers the broad H$\beta$ lines and ZTF $r$-band covers the broad H$\alpha$  lines for Mrk~1018 (z=0.043), 
that is the intrinsic optical color is contaminated by delayed broad Balmer lines and also by inter-band time delay, 
while the intrinsic mid-infrared color is just contaminated by the later. 
Therefore, the intrinsic optical and mid-infrared color$-$magnitude relations are derived by eliminating the potential contaminations 
(see Appendix~\ref{color} for more details), where the inset panel presents the light-curve profiles of the outburst phase 
that the potential contaminations are eliminated ($x$-axis is in units of JD-2,400,000 days, and $y$-axis is in units of magnitudes). 
In the bottommost row (panels ad and bd), 
we remove the underlying linear variation trend from the intrinsic color$-$magnitude relations given above (see Appendix~\ref{color} for more details), 
then calculate the net optical and mid-infrared color$-$magnitude relations that reflects the nature of the outburst phase. 
}
\label{figcolor}
\end{figure*}

\subsection{Multi-scale Outburst and Accretion Physics} \label{outburst}
During the observation period at LJT, we find that Mrk~1018 exhibited a significant outburst around 2020 (see Figure~\ref{longvari}). 
To explore multi-band variability, we gathered archival photometric data for optical and mid-infrared band as described above. The results are shown in the left panels of Figure~\ref{cle1} alongside the light curves of broad emission lines. 
The optical light curves include the ZTF $g$-, $r$-, and merged $g$\&$r$-bands and the mid-infrared light curves include WISE $W1$- and $W2$-bands. 
Multi-band light curves distinctly present the outburst feature of Mrk~1018 around 2020, rising abruptly within a timescale of about two hundred days from the long-term dimming state (see Figure~\ref{longvari}). 
In columns (5-7) of Table~\ref{tab1}, we present the typical values of the broad H$\alpha$ line fluxes and related parameters during the outburst, 
including the maximum values and those measured before and after but near the outburst. 
We find that the Eddington ratio changed more than 12 times throughout the outburst phase. 
In addition, the light curve of [O~{\sc iii}]$\lambda5007$, depicted in Figure~\ref{oiii} with a minor scatter of 6.9\% 
indicates that there is no variability in [O~{\sc iii}]$\lambda5007$. 
This finding suggests that the emissions from narrow-line region (NLR) are unaffected by the outburst event, 
or that the outburst did not introduce new components into the NLR. 

In order to quantify the characteristics of the outburst, we attempted to model the light-curve profiles of the outburst using various functions after detrending, 
identifying the Gaussian function as the most suitable model, the results are reported in panel (a) of Figure~\ref{cle1}. 
The details regarding the fitting processes and arguments refer to Section~\ref{rob} and Appendix~\ref{model} and \ref{color}. 
The typical timescales of multi-band outburst (e.g., $T$=$2\times2.35\sigma$) and peak times ($t_{\rm peak}$) 
were calculated from the modelings and are summarized in Table~\ref{tab2}. 
We find that the multi-scale outburst timescales exhibit the relation $T_{g}<T_{\rm H\beta}<T_{\rm H\alpha}<T_{W1}<T_{W2}$, 
while the peak times follow $t_{{\rm peak},g}<t_{\rm peak,H\beta}<t_{\rm peak,H\alpha}<t_{{\rm peak},W1}<t_{{\rm peak},W2}$, 
where the subscripts denote different bands or broad emission lines (see Figure~\ref{cle1}). 
These relations demonstrate that the multi-scale variabilities during the outbursts in Mrk 1018 are interrelated with different time delays 
and also suggest that the timescales and peak times increase with greater physical distance from the central SMBH, 
aligning with the echo phenomenon. 

The color$-$magnitude diagram is a widely used tool for examining AGN variability. 
Many efforts have explored the color$-$magnitude relation for CL-AGNs, 
revealing that their optical spectra exhibit a bluer-when-brighter behavior, 
while mid-infrared spectra show a redder-when-brighter trend (e.g., \citealt{Sheng2017,Yang2018,Graham2020}). 
Using Gaussian modelings, we have constructed the optical and mid-infrared color$-$magnitude diagrams for the outburst phase, 
as displayed in Figure~\ref{figcolor} (see  Appendix~\ref{color} for details). 
Indeed, the optical and mid-infrared variations show different trends in Mrk 1018. 

In the optical band, 
we observed a bluer-when-brighter trend, aligning with findings in normal AGNs. 
This implies that the mechanisms underlying extreme variability in CL-AGNs are likely analogous to 
those responsible for the stochastic variability observed in normal AGNs. 
The maximum variation in optical color reaches $\Delta_{\rm color}$=0.47~mag, alongside a significant flux increase by a factor of 17 for broad lines
(a proxy of the UV continuum), indicating that the nucleus spectrum becomes much bluer during outburst. 
However, the nucleus still has a low accretion rate even at the peak of the outburst ($L_{\rm bol}/L_{\rm Edd}\sim$0.012). 
It is important to note that although the rising phase of the outburst lacks data, 
the declining phase are well captured, with the results from this stage actually confirming these certain relations at least. 

Many models of variability could account for the extreme outburst, as discussed in details by \cite{Brogan2023}. 
In summary, a clumpy accretion disk could function on a viscous timescale, yet fails to match the outburst duration (about a year). 
Quasi-periodic eruptions can be linked to a warped accretion disk model, but we do not detect quasi-periodic outbursts during our monitoring period. 
Tidal Disruption Events (TDE), occurring when a star wanders close to a SMBH, can produce a short and dramatic outburst; 
however, the light-curve profiles of the outburst in Mrk 1018 cannot be fitted by a typical power-law model associated with TDEs (also see \citealt{Brogan2023}). Besides, during the dimming phase, the nucleus's spectrum become significantly reddens, in conflict with the color variations expected form TDEs 
(for additional arguments see Section~\ref{fullcycle}). 
The chaotic cold accretion of SMBHs, potentially linked to the viscous process, also fails to explain the outburst timescale. 
Our findings challenge all these models. 

It is generally accepted that the central engine of low-accretion/low-luminosity AGNs 
(e.g., those below a characteristic accretion rate of $\sim$1\% of Eddington ratio) 
consist of an inner hot advection-dominated accretion flow (ADAF, \citealt{Abramowicz1995}) 
and an outer truncated cold rotation-dominated thin disk (SSD, \citealt{Shakura1973,Ho2008}). 
The central engine of Mrk~1018 likely aligns with this two-component accretion disk model, as it overall stays below the 1\% Eddington limit 
during both the outburst and past 45 years of observations (see Table~\ref{tab1}). 
Based on this model, a radiation pressure instability may occur in a narrow transition region between the two components \citep{Sniegowska2020}, 
potentially leading to short-timescale outbursts and inducing changing-look behavior. 
This variability model actually applies in a viscous framework, predicting that the nucleus's spectrum should become significantly bluer, 
and its viscous timescale can be shortened by several factors (dependent on the ratio of the transition-region size to its distance 
from the central SMBH) compared to the viscous timescale of the Shakura-Sunyaev disk \citep{Shakura1973}. Remarkably, 
our findings closely align with the predictions of this variability model in light of the outburst timescale and nucleus spectrum. 
Another competing mechanism is the Bondi accretion of unstable surrounding gases during the transition between active and passive SMBHs \citep{Wang2024}. We find that the changing-look timescale observed for Mrk~1018 in this study can be explained by a sudden inflow of 
surrounding gases within the Bondi accretion radius, situated between the BLR and the NLR (see Equation 3 of \citealt{Wang2024}). 

In the mid-infrared band, we detected the color transitions occurred in the rising/dimming phases. Specifically, 
the mid-infrared spectrum shows an apparent redder-when-brighter relation when the flux is below approximately half of the maximum value, 
while it transitions to a bluer-when-brighter relation when the flux crosses this threshold (see panels~(ba) to (bd) in Figure~\ref{figcolor}). 
This contrasts with earlier findings showing a monotonically redder-when-brighter trend in the mid-infrared spectrum of CL-AGNs 
\citep{Sheng2017,Yang2018,Graham2020}. 
We noted that these previous results were derived  directly from the observed data, as illustrated in the panel (bb) of Figure~\ref{figcolor}, 
where the $W1$-$W2$~vs.~$W1$ relation represented by black circles demonstrated the redder-when-brighter relation, 
and two red circles from the high-luminosity state of the outburst phase appear to disrupt this monotonicity. 
If the mid-infrared emission at 3.4 and 4.6~$\mu$m arise from hot dust heated by AGNs activities (\citealt{Netzer2013}), 
the redder-when-brighter trend contradicts the viscous heating model of AGNs on the torus. 
Any star formation occurring within or outside the torus should heated the torus, 
resulting in a redder emission from the torus with increasing star formation rates, 
which aligns with the redder-when-brighter interpretation of the mid-infrared spectrum \citep{Yang2018}. 
However, this explanation does not account for the outburst timescale observed in Mrk~1018, 
as the stellar heating timescale on $W2$-band emission region should be comparable to the stellar evolution timescale. 
Our modeling of the outburst profiles indicates that the $W2$-band light-curve is broader than that of the $W1$-band, 
suggesting a more extended emission region or a smoother emissivity in the $W2$-band. 
Consequently, the rise in $W2$-band flux occurs earlier than that in the $W1$-band, while the dimming in $W2$ lags behind $W1$, 
resulting in the mid-infrared color exhibiting a redder-when-brighter trend during low-luminosity states. 
Interestingly, the mid-infrared bluer-when-brighter relation observed for the first time in relatively high-luminosity states, 
supports the notion that hot dust is heated by AGN activities \citep{Netzer2013}. 
This is further supported by the correlation between the mid-infrared outburst and nucleus outburst (see Figure ~\ref{cle1}). 

\begin{deluxetable}{ccc}
\tablecolumns{3}
\tabletypesize{\footnotesize}
\setlength{\tabcolsep}{4pt}
\tablewidth{4pt}
\tablecaption{Characteristics of the outburst \label{tab2}}
\tablehead{
\colhead{Band/Line}                                    &
\colhead{FWHM}                         &
\colhead{Peak ($t_{\rm peak}$)}                      \\
\colhead{} &
\colhead{(days)} &
\colhead{(JD-2,400,000 days)} 
}
\startdata
ZTF-$g\&r$       & 207$\pm$12 & 59049$\pm$7  \\
ZTF-$g$       & 186$\pm$12 & 59042$\pm$8  \\
ZTF-$r$       & 228$\pm$21 & 59055$\pm$12  \\
H$\beta$      & 192$\pm$5   & 59071$\pm$3  \\
H$\alpha$    & 206$\pm$7   & 59112$\pm$3  \\
WISE-$W1$ & 302$\pm$27 & 59140$\pm$8  \\
WISE-$W2$ & 345$\pm$19 & 59195$\pm$9 
\enddata
\end{deluxetable}

\subsection{Emission Region Sizes and SMBH Mass}  \label{sizeBHmass}  
During our spectroscopic monitoring of Mrk~1018, the availability of multi-band light curves enabled us to perform reverberation mapping 
measurements to determine the size of different emission regions in Mrk~1018. Utilizing the interpolated cross-correlation analysis  (\citealt{Sun2018}), 
we obtained the rest-frame time delays to be $32^{+36}_{-8}$ days for the broad H$\beta$ line, 
$83^{+23}_{-33}$days for the broad H$\alpha$ line, $78^{+57}_{-21}$ days for WISE $W1$-band, and 158$^{+56}_{-84}$ days for WISE $W2$-band, 
all relative to the the combined ZTF $g\&r$-band. 
The time delays between the mid-infrared band and ZTF $g\&r$ band were corrected for the cosmological time dilation 
and wavelength dependence using the total correction factor of (1+$z$)$^{\gamma}$-1=(1+$z$)$^{-0.24}$ (\citealt{Chen2023b}). 
Unfortunately, the seasonal gaps and sparse sampling result in relatively larger uncertainties in the obtained lags.
As a verification, from the Gaussian modeling of outburst profile 
(see Section~\ref{outburst}, \ref{rob} and Appendix~\ref{model} and \ref{color}), 
the time delays between different bands can be measured by the offsets in peak times. 
For the optical continuum variability, we test photometric light curves from  ZTF $g$-, $r$- and combined $g\&r$-band respectively and only find minor discrepancies in the typical timescales and peak times of the outburst, attributed to the influences of broad emission lines in the passbands.
If using the ZTF $g$-band as the reference, the rest-frame time delays is 28$\pm$10 days for the broad H$\beta$ line, 
67$\pm$10 days for the broad H$\alpha$ line, 97$\pm$16 days for WISE $W1$-band, and 152$\pm$17 days for WISE $W2$-band. 
It should be noted that the detrending windows used to subtract the underlying flux in the dimming state were chosen to be close to the outburst 
(JD=2458700 and 2459630 for optical bands and JD=2458600 and 2459990 for mid-infrared bands), 
however, we find that wider detrending windows do not influence the modeling outcomes. 
Considering the outburst profiles are effectively modeled by Gaussian functions (see Section~\ref{rob} and Appendix~\ref{model} and \ref{color}), 
we tentatively conclude that the above time-delay measurements provide reasonable estimates for the sizes of emission regions.

The virial mass of the SMBH in Mrk~1018 can be estimated using the recipe $M_{\bullet}=fc\tau v^{2}/G$, 
where $f$ represents the virial factor of the BLR, $c$ denotes the speed of light, $c\tau$ indicates the BLR radius measured through 
reverberation mapping method, $v$ refers to the rotation velocity of the BLR (characterized by the FWHM or velocity dispersion of broad lines), 
and $G$ is the gravitational constant. 
From the decomposed broad H$\beta$ lines during the outburst phase (from JD 2458811 to 2459433; see panel~(c) of Figure~\ref{cle1}), 
we obtain the average FWHM of broad H$\beta$ line to be 5475$\pm$1313~${\rm km\,s^{-1}}$ after correcting for the broadening effects 
(see Section~\ref{specAnalysis} for details on line-width measurements), with the uncertainty set by the standard deviation.  
Using the FWHM and time delay of the broad H$\beta$ line (28$\pm$10 days) and adopting  $f$=1, we estimate the virial mass of the SMBH in Mrk~1018 
(virial-based) to be $M_{\bullet}$=$(1.64\pm0.98)\times10^{8}M_{\sun}$. 

\begin{figure}[htb]
\centering
\includegraphics[angle=0,width=0.49\textwidth]{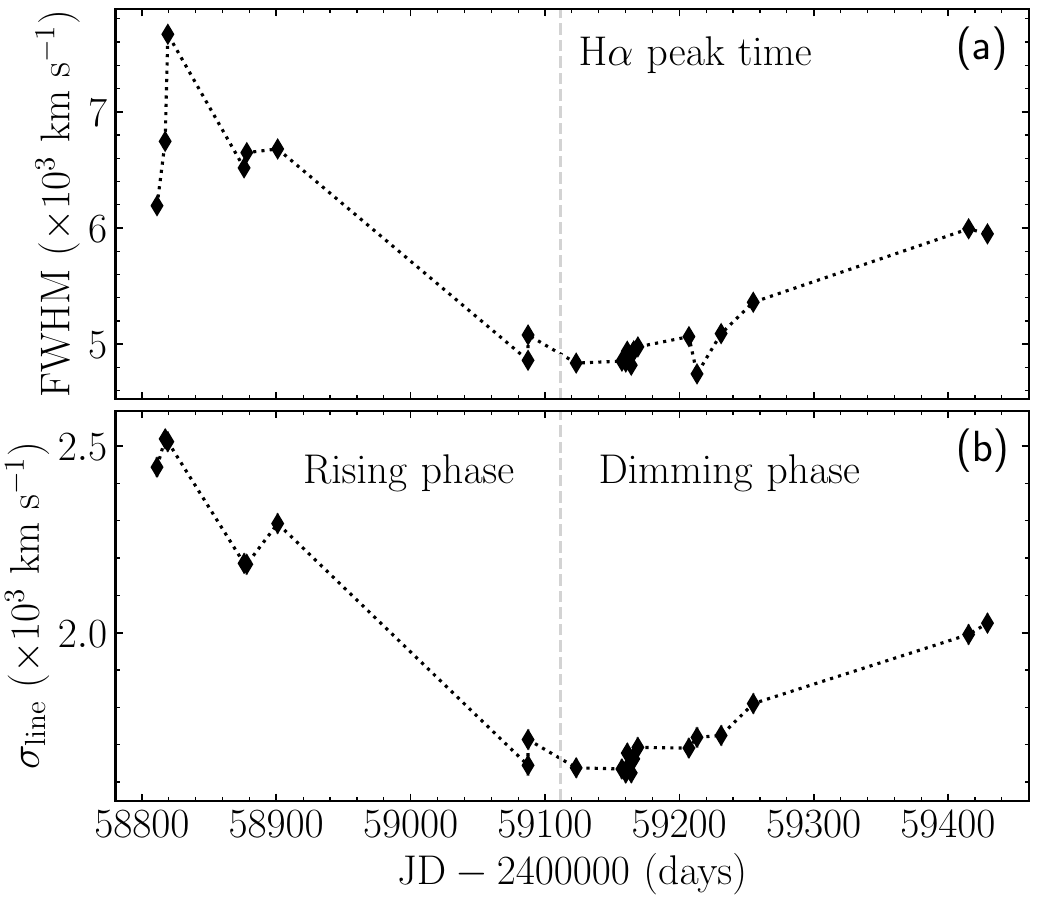}
\caption{
Evolution of the broad H$\alpha$ line width for Mrk~1018 during the outburst phase (JD2,458,811-2,459,433). 
The panel (a) is for the FWHM and the panel (b) is for $\sigma_{\rm line}$. 
The vertical dashed lines indicate the peak time of the broad H$\alpha$ line. 
The spectral broadening in each observation, evaluated by comparing line widths of [O~{\sc iii}]~$\lambda$5007 from SDSS and LJT spectra, 
has been corrected (see Section~\ref{specAnalysis}). 
We use the line-width scatter of [O~{\sc iii}]~$\lambda$5007 as the systematic uncertainty of the broad-line width. 
}
\label{OBlinewidth}
\end{figure}

The virial SMBH mass relies on the assumption that the BLR is virialized. 
However, in cases with rapid variability or outbursts, the BLR may not be virialized, 
leading to unreliable virial mass estimates from broad emission lines. 
For instance, during the changing-look transition of the true type 2 AGN 1ES~1927+654 over months, 
\cite{Li2022} found that the time lags of the broad emission lines relative to the optical continuum significantly exceed the radius-luminosity relations (\citealt{Bentz2013}), and the single-spectrum virial masses (\citealt{Greene2005}) were considerably higher than independent measurements using the relation between SMBH masses and the properties of host galaxies. Based on various observational evidence, 
including a decrease in both continuum flux and line width (which is often seen in TDEs; \citealt{Holoien2016,Leloudas2019}), 
\cite{Li2022} proposed that the SMBH in 1ES~1927+654 had captured a surrounding star (as in TDEs) resulting in a super-Eddington/high accretion system and suggested that its BLR was in the process of formation and rapid evolution, likely not yet virialized. 
Similarly, the spectral phenomena observed in Mrk~1018 (as discussed later) resemble those seen in 1ES~1927+654 (\citealt{Li2022}). 
This motivates us to examine the evolution of the broad H$\alpha$ line width (FWHM and $\sigma_{\rm line}$). 
As shown in Figure~\ref{OBlinewidth}, we find that the variations in line widths during the outburst phase align with the normal breathing effect of the BLR, 
indicating that increased continuum radiation enhances the broad-line emissivity of BLR gases at greater distances, 
resulting in a decrease in line width with rising continuum emission (and vice versa). 
This analysis implies that the nucleus outburst of Mrk~1018 did not significantly alter the kinematics of its BLR. 

To further verify the virial mass, we utilize the properties of the host galaxy to independently estimate the SMBH mass 
using the $M_{\bullet}-\sigma_{*}$ relation, where $\sigma_{*}$ represents the stellar velocity dispersion; \citealt{Gebhardt2000,Ferrarese2000}). 
Since no stellar velocity dispersion is available for the bulge of Mrk~1018, 
we approximate the stellar velocity dispersion by the velocity dispersion of the ionized gas, which is derived from the narrow forbidden emission lines.  
It has been shown to be approximately valid for a large variety of AGNs (e.g., \citealt{Nelson1996,Ho2009,Kong2018}). 
We find $\sigma_{\rm gas}$=176~${\rm km\,s^{-1}}$ for the [O~{\sc iii}]$\lambda$5007 line from the SDSS spectrum observed in 2000, 
which provides the highest resolution among all spectra available for Mrk~1018. 
According to the latest $M_{\bullet}-\sigma_{*}$ relation for early-type galaxies \citep{Greene2020}, 
we derive $M_{\bullet}=(1.61\pm0.41)\times 10^{8}~M_{\sun}$, remarkably consistent with virial-based SMBH mass. 
This results in the Eddington luminosity $(2.03\pm0.52)\times10^{46}$~erg~s$^{-1}$. 
By combining with the derived bolometric luminosity (see Section~\ref{dataanalysis}),  
we compute the Eddington ratio for each observation, as shown in Figure~\ref{longvari}. 
Over the past 45 years, the Eddington ratio has varied from 2.0$\times10^{-5}$ to 0.02 , 
suggesting that Mrk~1018 has been powered by very low accretion rates. 

\subsection{Full-cycle changing-look transition and type transition}  \label{fullcycle}
It is generally accepted that a full-cycle changing-look transition should last several decades, which may be linked to the 
viscous heating timescale of the accretion disk (e.g., \citealt{Shakura1973,Runnoe2016}). 
However, Our observations show that Mrk~1018 experienced a full-cycle changing-look transition within just one year following a multi-scale outburst. 
Its type changed from Type 1.8 to Type 1.5/1.2 and then returned to Type 1.8 over the course of the outburst (see panel~(b) of Figure~\ref{cle1}). 
Notably, the broad H$\beta$ and H$\alpha$ lines were nearly absent before the outburst, appeared significantly during the outburst, and then almost vanished 
by the end of the outburst (see panel~(c) of Figure~\ref{cle1}). 
The short appearance and rapid disappearance of broad Balmer lines in Mrk~1018 exhibit significant discrepancies with theoretical predictions. 

By integrating the archival spectral data, we find that Mrk~1018 has experienced a full-cycle type transition, encompassing 
sub-types 1.0, 1.2, 1.5, 1.8/1.9 in a continuous manner (see Figure~\ref{longvari} and \ref{belevo}). 
Our observation from LJT effectively bridge the gap between Type 1.2 and Type 1.8/1.9. 
We investigate the type transition of Mrk~1018 in relation to the Eddington ratio, as shown in panel~(a) of Figure~\ref{blrevo}, 
and find a positive correlation between the type transition and Eddington ratio. 
This shows that strongly varying accretion rate are responsible for regulating the AGN type transitions. 
Moreover, Mrk~1018 actually represents the first full-cycle observational case, revealing that the intrinsic broad-line emission of AGNs population follows 
an evolutionary sequence: Type 2-Type1.9/1.8-Type1.5/1.2-Type1 as the accretion rate increases \citep{Stern2012a,Stern2012b,Elitzur2014}. 

As demonstrated in 1ES~1927+654 (\citealt{Trakhtenbrot2019,Li2022}), TDEs can act as a trigger for changing-look behavior in AGNs. 
However, in the case of Mrk~1018, the changing-look transition that took place around 2020 was not induced by a TDE. 
One reason, as mentioned in Section~\ref{sizeBHmass}, is that the evolution of broad line width during the dimming phase 
(normal breathing, see Figure~\ref{OBlinewidth}) contradicts the TDE scenario 
(typically, the line width decreases with time, while the continuum flux drops; \citealt{Holoien2016,Leloudas2019}). 
Section~\ref{outburst} and Appendix~(\ref{model} and \ref{color}) present additional reasons. 
Thus, changing obscuration and changing state remain potential origins for the changing-look behavior of Mrk~1018. 
The correlations of multi-scale variability effectively rule out the former case since these correlations would be significantly weakened 
by changing in the covering factor of obstructions along the line of sight (also see \citealt{Sheng2017}). 
Furthermore, no ultraviolet and X-ray absorbers have been detected in nucleus of Mrk~1018 over the past decades \citep{ Krumpe2017,Brogan2023}, 
which also disfavors the changing obscuration hypothesis. 
In conclusion, the changing accretion-driven radiation field (or changing-state) might serve as the changing-look mechanism of Mrk~1018, 
supported by observational evidence of an accretion-dependent full-cycle type transition (see panel~(a) of Figure~\ref{blrevo}). 

\begin{figure*}[htb]
\centering
\includegraphics[angle=0,width=0.49\textwidth]{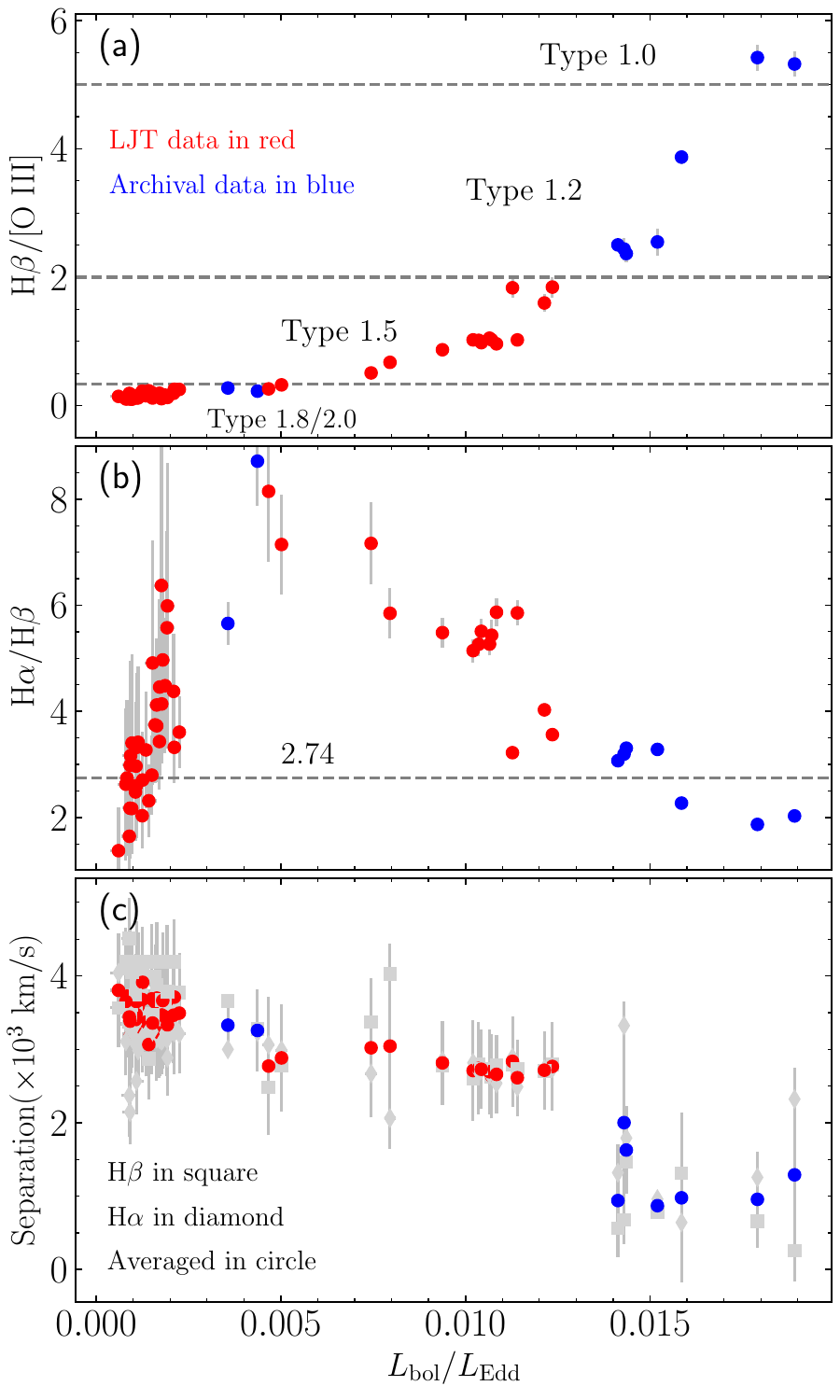}
\includegraphics[angle=0,width=0.49\textwidth]{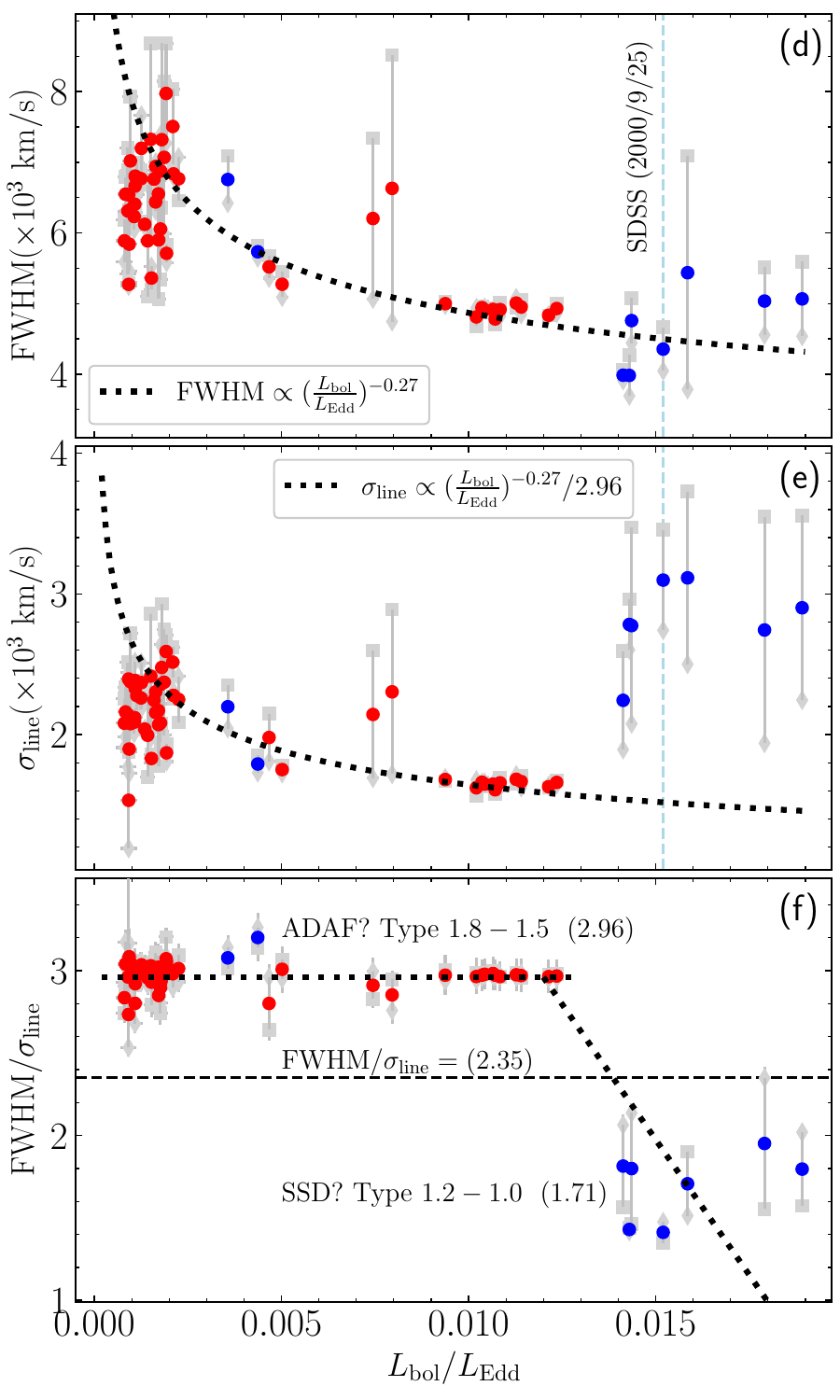}
\caption{
Accretion-dependent type transitions and the broad-line properties in Mrk~1018. 
The data observed by LJT is shown by red points and the archival data is shown by blue points. 
Panel~(a-c) show the changes of the AGNs type (H$\beta$/[O~{\sc iii}]$\lambda5007$), the Balmer decrement (H$\alpha$/H$\beta$), and 
the peak separation of the double-Gaussian broad-line models with the Eddington ration, respectively. 
In panel (b), the grey dashed line presents the Baker–Menzel Case B value of 2.74 \citep{Osterbrock2006}. 
In panel (c), the H$\beta$ and H$\alpha$ peak separations are represented by squares and diamonds, respectively, and the average peak separation is shown as points. 
Panel~(d-f) shows changes of the broad H$\beta$ FWHM,  the H$\beta$ $\sigma_{\rm line}$ (velocity dispersion), and the line width ratio FWHM/$\sigma_{\rm line}$ with the Eddington ratio, respectively. 
In panel~(f), Type 1.8-1.5 and Type 1.2-1.0 phases, potentially linked to different accretion modes (ADAF and SSD), 
exhibit distinct FWHM/$\sigma_{\rm line}$. 
Across panels panel (a-f), to ensure reliable measurements on the parameters, particularly for those involving broad line signals, 
we select the spectra where the H$\beta$/[O~{\sc iii}]$\lambda$5007 ratio exceeds 0.1. 
The spectral broadening in each observation 
(determined by comparing the line width of [O~{\sc iii}]$\lambda$5007 in the SDSS and LJT spectra) 
has been corrected (see Section~\ref{specAnalysis}). 
}
\label{blrevo}
\end{figure*}

\subsection{Broad-line Properties and BLR Physics}  \label{blrph} 
Over the past 45 years, Mrk~1018 has shown significant variability in its broad lines (Figures~\ref{longvari} and \ref{belevo}), 
providing a unique perspective to study the BLR. 
Our research reveals that the Balmer decrements have varied significantly from about 1.5 to 9. 
By analyzing the dependence of the Balmer decrement variations on Eddington ratio, as shown in panel~(b) of Figure~\ref{blrevo}, 
we find for the first time that the Balmer decrement initially increases and then decreases with increasing Eddington ratio, 
with a turnover occurring at an Eddington ratio around $\sim$0.004, which roughly corresponds to the transition point between Type 1.8 and Type 1.5. 
In essence, the Balmer decrement increases to the maximum and then decreases again, 
a comparable phenomenon has likely been observed in other CL-AGNs (see Figure 13 of \citealt{Panda2024}). 
In Sections~\ref{fullcycle}, we argued that the changing obscuration is not the mechanism for the changing-look/type transition in Mrk~1018.
Thus, the Balmer decrement serves as an indicator for the ionization effects to the BLR rather than the internal reddening. 
Previous studies have identified an anti-correlation between Balmer decrement and accretion rate 
in individual AGNs or normal AGN samples (e.g.,  \citealt{Wu2023,Ma2023}). 
In the case that there is no nucleus absorption, this anti-correlation can be attributed to the optical depth effects 
as mentioned in previous studies (e.g., \citealt{Netzer1975,Korista2004,Wu2023}). 
This relationship arises because the optical depths of Balmer lines within in the BLR of AGNs are proportional to the ionizing luminosity, 
where the optical depth in H$\alpha$ is larger than H$\beta$ for a given ionizing luminosity. 
Interestingly, we have observed for the first time that in Mrk~1018, 
the Balmer decrement increases to a peak and then decreases (refer to panel~(b) of Figure~\ref{blrevo}). 
The early theoretical calculations suggested that the variation of Balmer decrement (H$\alpha$/H$\beta$) 
in dense environment of AGNs is dependent on the optical depth and exists a turnover, potentially occurring in a range of 
$60\le \tau_{\rm H\alpha} \le 120$ (see \citealt{Netzer1975}), where $\tau_{\rm H\alpha}$ denotes the H$\alpha$ optical depth. 
Out findings provide direct observational support for these theoretical predictions. 

Our study shows that the broad H$\beta$ and H$\alpha$ line profiles have significantly changed, 
evolving from single-peaked profiles in Type 1.0-1.2 phase to double-peaked profiles in Type 1.5-1.8/1.9 phase (see Figure~\ref{belevo}). 
Double-peaked profiles are a hallmark feature of broad Balmer lines during the changing-look transition observed at LJT (in Type 1.5), 
and the broad H$\beta$ and H$\alpha$ profiles, which exclude the narrow lines, clearly display these features (see Figure~\ref{specfit}). 
Previous studies had shown that those AGNs classified as Type 1.8/1.9, which typically have low-Eddington ratio, 
frequently display double-peaked broad-line profiles, whereas the majority of Type 1.0/1.2 AGNs do not 
(e.g., \citealt{Ho2000,Ho2008,Strateva2003,Elitzur2014}). 
Interestingly, the changes in broad-line profiles of Mrk~1018 is consistent with findings from large-sample analyses. 
In Section~\ref{specAnalysis}, we show that all broad H$\beta$ and H$\alpha$ lines observed in the past 45 years 
can be effectively modeled with two Gaussians (see Figure~\ref{specfit}, also see Figure 3 of \citealt{Kim2018} for the early spectra). 
This allows us to derive key parameters of the broad-line characteristics, 
such as peak separation and broadening corrected line width of FWHM and $\sigma_{\rm line}$. 
It is known that the shape of broad-line profile can be parameterized by the line-width ratio of FWHM to line dispersion 
(i.e., FWHM/$\sigma_{\rm line}$, see \citealt{Kollatschny2011,Kollatschny2013}). 
A Gaussian profile has a FWHM/$\sigma_{\rm line}$ of 2.35, a rectangular profile has a ratio of 3.46, 
and a Lorentzian profile shows FWHM/$\sigma_{\rm line}<2.35$ (\citealt{Kollatschny2011}). 
Below we explore the BLR properties for Mrk 1018 using these parameters. 

The parameters vary by nearly a factor of two and are plotted with Eddington ratio in Figure~\ref{blrevo}. 
For the first time, we find that the peak separation decreases with increasing Eddington ratio, 
with a maximum difference of 2500~km/s (panel (c) of Figure~\ref{blrevo}). 
Overall, the FWHM appears to exhibit an exponential decrease as the Eddington ratio rises (panel (d) of Figure~\ref{blrevo}). 
Using the virial relation of FWHM$\propto R_{\rm BLR}^{-0.5}$ and empirical relation of $R_{\rm BLR}\propto L_{\rm 5100}^{0.533}$ (\citealt{Bentz2013}), 
we derive FWHM$\propto L_{\rm 5100}^{-0.27}$, which accordingly leads to FWHM$\propto (L_{\rm bol}/{L_{\rm Edd}})^{-0.27}$. 
The dotted curve in panel (d) of of Figure~\ref{blrevo}, along with the observed data, represent this relation. This suggests that the BLR kinematics in Mrk~1018 is consistent with the virial motion  
and the global kinematics of the BLR is stable (at least in the low-accretion system as Mrk~1018). 
In practice, the variations in peak separation and FWHM can be attributed to the well-known breathing effect of the BLR. 
Specifically, an increase in ionizing luminosity from a higher accretion rate enhances the broad-line emissivity of the BLR 
at larger distances, resulting in a decrease of peak separation along with the decrease in the rotation velocity of the BLR. 
In addition, the $\sigma_{\rm line}$ also shows a similar trend as the FWHM (see the dotted line) 
when the Eddington ratio is below $\sim$0.013 (corresponding to Type 1.8-1.5 phase), but it is subject to a sharp rise 
when the Eddington ratio exceeds 0.013 (corresponding to Type 1.2-1.0 phase; see panel (d) of Figure~\ref{blrevo}). 
This implies that a simple breathing effect of the BLR can not fully account for the changes of $\sigma_{\rm line}$. 

In order to understand the aforementioned differences in the evolution of FWHM and $\sigma_{\rm line}$, 
we turn next to examining the shape of broad line profile, as shown in panel (f) of Figure~\ref{blrevo}. 
When the nucleus accretion is below $\sim$0.013 times Eddington ratio (in the Type 1.8-1.5 phase), 
suggesting the nucleus is likely powered by ADAF (\citealt{Abramowicz1995}), 
FWHM/$\sigma_{\rm line}$ remains constant with an average value of 2.96. 
This indicates that that the broad-line profile approximately resembles a rectangular shape. 
Conversely, when the nucleus accretion exceeded roughly $\sim$0.013 times Eddington ratio (in the Type 1.2-1.0 phase), 
indicating a probable transition to SSD (\citealt{Shakura1973}), FWHM/$\sigma_{\rm line}$ sharply drops to an average of 1.71. 
This suggests that the broad-line profile tends to be a Lorentzian shape. 
It had been proposed that the dominated broadening mechanism for the broad line profiles arises from Keplerian rotation 
(implying FWHM$\propto R_{\rm BLR}^{-0.5}$ for the virialized BLR), 
and a Lorentzian shape for the broad line (with FWHM/$\sigma_{\rm line}<2.35$) is attributed to turbulences in the BLR 
(\citealt{Osterbrock1978,Kollatschny2011,Kollatschny2013}). 
Consequently, as turbulence in the BLR increases, FWHM/$\sigma_{\rm line}$ decreases. 
As mentioned above, the accretion flow in Mrk~1018 may transition from ADAF to SSD (also see \citealt{Lyu2021}), 
where an optically thick geometrically thin disk extends down to the innermost stable circular orbit (\citealt{Capellupo2015}), 
resulting in an extended accretion disk. In this scenario, turbulent motions in the outer accretion disk produce a Lorentzian profile, 
as suggested by several studies (e.g., \citealt{Veron2001,Sulentic2002,Goad2012}). 

There are two popular models that can explain the double-peaked nature of broad emission lines. 
For the first one, based on the failed radiatively accelerated dusty outflow model (FRADO, see Figure 1 of \citealt{Czerny2011,Naddaf2022}), 
some simulations have shown that the low-ionization broad lines usually present double-peaked profiles when nucleus operates 
at a few percent of Eddington ratio and transition to single-peaked profiles at high Eddington ratio 
(\citealt{Naddaf2022}; Wu et al. 2024, submitted). Also, as accretion rate increases, the peak separation decreases. 
This prediction actually is consistent with our observational finding regarding the peak separation (see panel~{\bf c} of Figure~\ref{blrevo}). 
The second  model is the Keplerian disk-like model, which also accounts for the occurrence of double-peaked emission lines \citep{Chen1989,Strateva2003}. 
In practice, these two models may be reconciled by considering an accretion-dependent BLR structure, 
which has been observed during full-cycle type transition of Mrk~1018 (see Figure~\ref{blrevo}) and in samples of AGNs (see \citealt{Elitzur2014}). 
It is plausible that when the effective temperature of the accretion disk falls below 1000~K, 
the dusty wind (comprising both BLR gas and dust) resides at the disk's surface (disk-like BLR) 
before being expelled from the disk and will rise to higher latitudes (bowl-like or spherical BLR) as the temperature increases. 
This transition results in a change from double-peaked to single-peaked profiles in the broad emission lines. 
Moreover, as the dusty wind is accelerated to higher latitudes, its turbulence likely increase dramatically, 
causing a sharp decrease in FWHM/$\sigma_{\rm line}$. 
Our observations in Mrk~1018 lend a support to the dusty-wind model. 

In summary, these findings suggest that a virialized BLR together with turbulent motion regulated by the accretion rate or state transition of accretion disk 
could account for various BLR phenomena observed in AGNs population. 
Alternatively, some reverberation mapping studies pointed to two distinct components of BLRs (\citealt{Hu2020,Nagoshi2024}). 
For instance, based on reverberation mapping of an extremely variable quasar (SDSS J125809.31+351943.0, J1258), 
\cite{Nagoshi2024} proposed that the BLR comprises of two distinct emission regions associated with the dusty torus (see Figure 11 of \citealt{Nagoshi2024}). 
One region, situated close to the SMBH and potentially at the surface of the accretion disk (as disk-like geometry), is likely formed from accretion related to the dust torus and produce the double-peaked profile. 
The other region, located relatively father from the SMBH and possibly at high latitude of accretion disk (as bowl-like geometry), 
is likely formed by sublimated dust torus, resulting in the single-peaked profile. 
\cite{Nagoshi2024} further claimed that the broad line emissions of disk-like region might occur independently of the nucleus's activity. 
However, in the case of Mrk~1018, we observe a strong correlation between the broad line emissions and the varying optical continuum 
during the double-peaked phase, in conflict with the above scenario of the two-component model. 

\section{Discussion} \label{dis}
\subsection{Comparison with the CLE in 1ES~1927+654} \label{comp} 
Similar to Mrk~1018, the previously recognized true type 2 AGN 1ES 1927+654 (\citealt{Panessa2002,Tran2011}) underwent an outburst in optical and UV by 4 magnitudes on December 23, 2017. 
This event, marked by the appearance of broad emission lines after about 100 days, is referred to as a changing-look phenomenon (\citealt{Trakhtenbrot2019}). 
Studies have examined the nature of 1ES 1927+654 (\citealt{Trakhtenbrot2019,Li2022,Laha2022}), 
focusing on the feeding mechanisms of the central engine and the formation of the broad line region (BLR). 
In brief, the optical continuum rises rapidly and then declines following $t^{-5/3}$ (\citealt{Trakhtenbrot2019}), 
a behavior consistent with the fallback accretion rate expected in TDEs. 
The significant 100-day delay in broad emission line formation suggests that the BLR emerges after the outburst and likely has not yet been virialized. 
The subsequent rapid secular evolution of the BLR kinematics results in significant changes in the broad line profiles, 
including variations in the Balmer decrements, profile shapes and line widths (\citealt{Li2022}). 
Interestingly, as the optical continuum flux decreases, the width of the broad lines is also diminished, 
a pattern commonly observed in TDEs and attributed to a decelerating outflow (\citealt{Holoien2016,Leloudas2019}). 
Moreover, the strength of the narrow Balmer lines systematically increases by a factor of 4. 
In conclusion, the outburst behavior in the optical continuum and overall evolution of the broad emission lines suggest 
that the CLE in 1ES 1927+654 is associated with a TDE (\citealt{Li2022}). 

Additionally, various outburst events have been detected in different AGNs over the past few decades (e.g., \citealt{Graham2020,Zhang2023}). 
For instance, the narrow-line Seyfert 1 galaxy CSS~J102913+404220 experienced multi-band outbursts around 2010, 
possibly also linked to a tidal disruption event (TDE). \cite{Zhang2022} found that its UV/optical continuum remained relatively constant over time, 
while the broad H$\alpha$ line increased in flux approximately a decade after the nuclear optical outburst. 
They suggested that this phenomenon might be due to the replenishment of gas and excitation in the BLR, 
potentially resulting from interactions with outflowing stellar debris.

At first glance, Mrk~1018 and 1ES~1927+654 appear to share similar phenomena, 
such as an outburst in optical continuum associated with the strong changes in their accretion rates, 
which illuminated the BLR and induced a full-cycle changing-look transition. However, the outburst characteristics 
suggest that the drastic change in the accretion rate of 1ES~1927+654 is attributed to a TDE, 
whereas in Mrk~1018, the outburst is most likely linked to the behavior of the accretion disk itself (for further details, see Section~\ref{outburst}). 
That is to say, the full-cycle changing-look transitions in each AGN might result from different feeding mechanisms. 
Furthermore, the evolution of broad emission lines also show differences between the two sources. 
Specifically, the Balmer decrement of the broad line (H$\alpha$/H$\beta$) in 1ES 1927+654 steadily increases from 1 to 7 following the emergence of the broad lines, 
while in Mrk~1018, the Balmer decrement  decreases from 9 to 1 as the Eddington ratio increases. 
Additionally, in 1ES~1927+654, the FWHM decreases with optical continuum flux, a trend commonly seen in TDEs, 
whereas in Mrk~1018, the FWHM displays an opposite trend. 
Moreover, \cite{Trakhtenbrot2019} and \cite{Li2022} revealed that the emergence of the broad Balmer line, 
significantly lags behind the rising optical continuum, 
suggesting a newly formed BLR. In contrast, Mrk~1018 showed weak broad H$\beta$ and H$\alpha$ lines 
(H$\beta$/[O~{\sc iii}]$\lambda$5007=0.1) before the outburst/changing-look event, 
and the variation of broad H$\beta$ line follows the optical continuum changes without significant deviation 
from the empirical BLR radius-luminosity relation. 
These differences imply different physical processes: 1ES~1927+654 may elucidate the formation of the BLR (see \citealt{Li2022}), 
while Mrk~1018 may illustrate the BLR evolution during AGN turn-on or turn-off. 

\begin{figure}[htb]
\centering
\includegraphics[angle=0,width=0.48\textwidth]{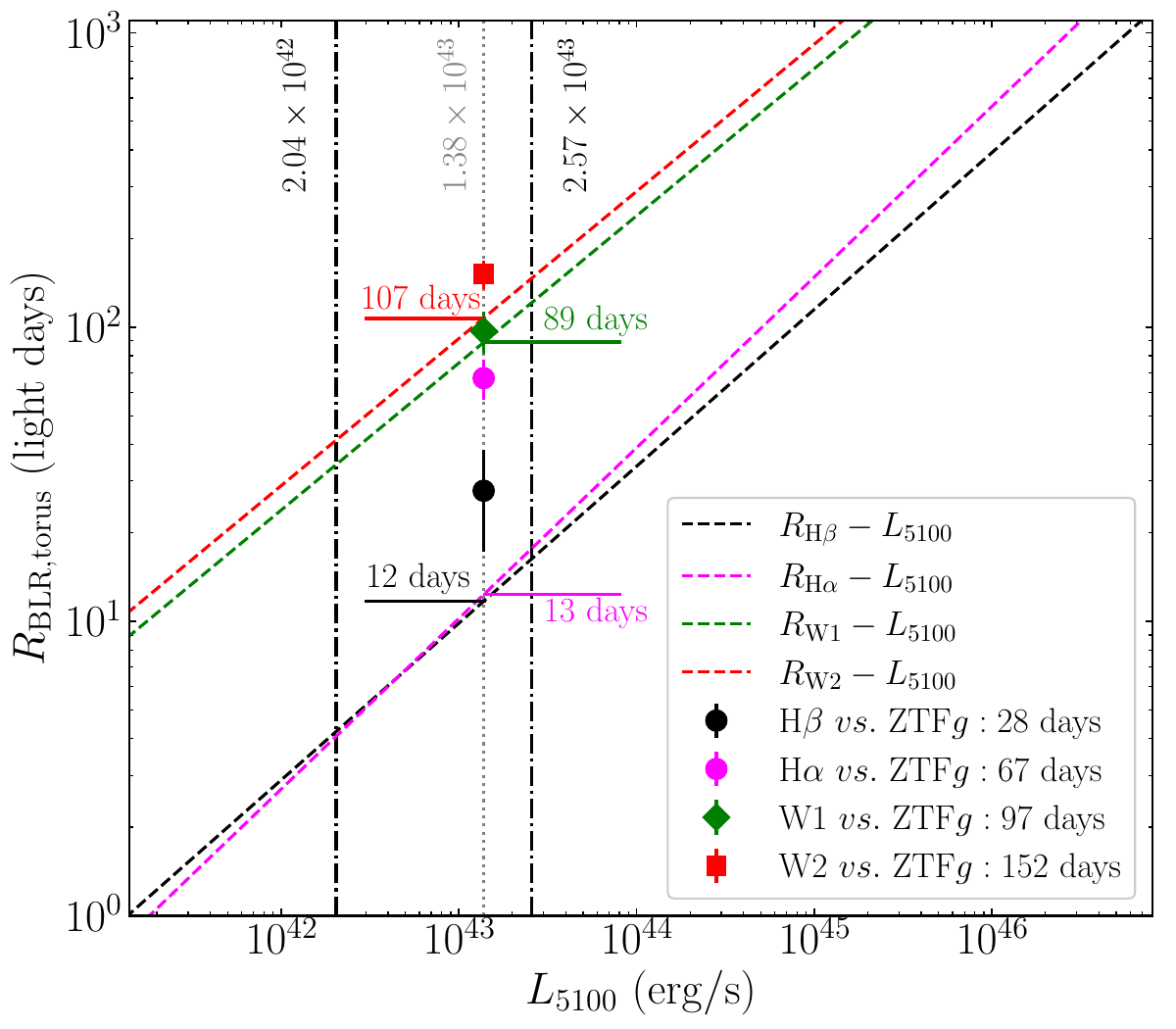}
\caption{
Relations between emission region sizes and optical luminosities. 
The dashed lines represent the empirical radius-luminosity relations for 
the H$\beta$ (in black) and H$\alpha$ (in pink) BLR (\citealt{Bentz2013,Cho2023}) and the torus (\citealt{Chen2023b}, in red and green). 
The colored horizontal line segments mark the sizes obtained from the corresponding empirical radius-luminosity relation (with the same color code).
The vertical black dot-dashed lines denote the minimum and maximum luminosities observed during the outburst phase, 
while the averaged luminosity is denoted by a gray vertical dotted line. 
The data points represent the estimated sizes of the BLR and torus estimated from the outburst light-curve modeling. 
}
\label{size}
\end{figure}

\subsection{Torus-BLR Connection}  \label{blrtorus} 
Over the past decades, great efforts have been made to establish the empirical radius-luminosity relations for selected AGN samples 
as shown in Figure~\ref{size} (\citealt{Bentz2013,Cho2023,Chen2023b}). 
Such as, the H$\beta$- and H$\alpha$-based BLR radius-luminosity relation 
$R_{\rm H\beta}=33.6~l_{44}^{0.533}$ \citep{Bentz2013} and $R_{\rm H\alpha}=38.9~l_{44}^{0.58}$ light days \citep{Cho2023}; 
the hot torus $W1$- and $W2$-band radius-luminosity relations
$R_{W1}=239~l_{44}^{0.5}$ and $R_{W2}=289~l_{44}^{0.5}$ light days \citep{Chen2023b}, 
where $l_{44}=L_{\rm 5100}/10^{44}~{\rm erg~s^{-1}}$ with $L_{\rm 5100}$ being the optical Luminosity at 5100~\AA. 
These relations, with a typical intrinsic scatter of $\sim$0.2~dex, provide an effective way to estimate sizes of BLR and 
torus for a given optical luminosity. 

For the average optical luminosity of 1.38$\times$10$^{43}$~erg~s$^{-1}$ (see Table~\ref{tab1}) during the outburst phase of Mrk~1018, 
we derive the expected sizes of the broad H$\beta$ and H$\alpha$ lines, $W1$- and $W2$-band emission regions (shown in Figure~\ref{size}), 
corresponding to the light travel times of 12~days, 13~days, 89~days and 107~days, respectively. 
During this phase, the torus size of Mrk~1018 is consistent with the empirical torus radius$-$luminosity relation if taking into account the scatter, 
whereas the sizes of broad H$\beta$ and H$\alpha$ BLRs are large than the expected sizes by 0.37 and 0.75 dex, respectively. In particular, 
it seems that the broad H$\alpha$ BLR is close to the mid-infrared emission region. 
Recent reverberation mapping studies found that the broad H$\beta$ line lags are shorter than predicted from the canonical BLR radius-luminosity relation for super-Eddington AGNs (\citealt{Du2016,Du2018}) and sub-Eddington AGNs (\citealt{Grier2017,Malik2024}). 
In contrast, in Mrk~1018, the BLR size observed during the outburst phase in turn is larger than predicted. 

Three factors may explain the large BLR radius observed during the outburst phase. 
First, the rapid and extreme outburst can drive disk winds, 
creating shocks in the dense BLR and pushing it away from the central SMBH. 
If we assume that the 16 light-day deviation of the broad H$\beta$ emitter from the radius-luminosity relation (\citealt{Bentz2013}) is due to this phenomenon over a rising timescale of about two hundred days, it would require a significant radial velocity of $2.4\times10^{4}$~${\rm km~s^{-1}}$, which is 4 times higher than the BLR's rotational velocity of 5475~${\rm km~s^{-1}}$. 
However, the virialized nature of the BLR (as shown in Figure~\ref{OBlinewidth} and panel (d) of Figure~\ref{blrevo}) contradicts this possibility. 
Second, historical observation shows that Mrk~1018 reverted from type 1.0 to type 1.8 around 2015, 
suggesting that the inner BLR gases may have been depleted prior to this transition (e.g., due to accretion). 
Consequently, the recent outburst could ionize gases located farther from the SMBH and closer to the dust torus's sublimation region, 
leading to the formation of low-ionization lines. However, the connection between the accretion flow and BLR gases remains unclear. 
The third possibility might be that the BLR is coupled with the dust torus, as proposed by various BLR models. 
For example, a tidal disruption of dusty clumps by the SMBH could form a BLR \citep{Wang2017}; 
The BLR model depicts two distinct emission regions associated with the dust torus (see Figure 11 of \citealt{Nagoshi2024}); 
and strong local dusty winds from the disk (i.e., FRADO, see Section~\ref{blrph}) could also contribute to low-ionization emission lines \citep{Czerny2011}. 
If the latter scenario is valid, the dusty wind, comprising of low-ionization BLR gas and dust, could generate mid-infrared photons, 
producing low-ionization broad lines near the torus. 

\subsection{Robustness of Light-curve Modeling and Time Delay Measurement} \label{rob} 
After detrending, we fitted all outburst light-curve profiles using Gaussian and GE models (see Appendix~\ref{model}), 
and examined the intrinsic optical and mid-infrared color$-$magnitude relations during the outburst (Appendix~\ref{color}). 
Although the rising phase of the outburst was not well sampled, 
several factors indicate that the Gaussian modeling of these profiles is both sensible and robust. 
The mid-infrared outburst profiles appear symmetric. Gaussian fitting for all outburst profiles yields lower 
Bayesian Information Criterion (BIC) and Akaike’s Information Criterion (AIC) values compared to other models, despite similar reduced $\chi^{2}$ values. 
Although the BLR radius from our light-curve modeling exceeds the estimate from the empirical BLR radius-luminosity relation, 
the obtained torus radius is well consistent with the empirical torus radius-luminosity relation (see Figure~\ref{size}). 
By assuming that the torus is unaffected by the outburst, this suggests that light-curve modeling in mid-infrared and optical bands 
provides reliable measurement of torus radius. 
Possible explanations for the larger BLR radius are discussed in Section~\ref{blrtorus}. 
We also found that the color$-$magnitude relations for the rising and dimming phases are consistent (Figure~\ref{figcolor}) 
when analyzed with Gaussian modeling data (the TDE model was excluded, see Section~\ref{fullcycle} and Appendix~\ref{color}), 
indicating that these outburst light-curve profiles are well-represented by Gaussian functions. 
Furthermore, the corrections of time delay on different bands (derived from Gaussian modeling) are likely reasonable since 
unreliable time-delay corrections would lead to large dispersion in these relations, such as different intercepts. 
In addition, if the different models are valid for describing the rising and dimming variability-behaviors, 
such the GE model considered in Appendix~\ref{color}, 
it leads to intricate color$-$magnitude relations as shown in Figure~\ref{figsp2}, 
characterized by different amplitudes, intercepts, and slopes. 
However, the underlying reasons for these complex relationships remain unclear. We anticipate that upcoming surveys, like LSST and SVOM (\citealt{Czerny2023,Xu2024}), will uncover more extreme events similar to those seen in Mrk~1018, facilitating a more thorough exploration of nuclear variability. 
 
\section{Summary} \label{sum}
We performed long-term spectroscopic monitoring of Mrk 1018 using Lijiang~2.4~m telescope from 2017 to 2024 to investigate the evolution of AGNs. 
Our observations revealed a full-cycle changing-look transition, triggered by a significant outburst from the nucleus within one year. 
Combining with the archival spectral data, we detected a full-cycle type transition in Mrk 1018 for the first time, 
encompassing Types 1.0, 1.2, 1.5, 1.8, 1.9 and 2.0, and found strong evidence that the full-cycle type transition is regulated by accretion. 
Our extensive investigation obtains the following findings: 

\begin{enumerate}
\item
During the full-cycle changing-look transition,  the luminosity of Mrk~1018 reaches a few percent of the Eddington limit (0.1$\sim$1\%), 
and the optical broad emission line exhibits a double-peaked shape. 
The multi-band light curves of the outburst, with a characteristic timescale of about one year, were found to be strongly correlated. The nucleus spectrum becomes bluer during the outburst. 
These characteristics suggest that the nucleus outburst is likely due to an instability in the accretion disk occurring in the 
transition region between the outer standard thin disk and an inner advection-dominated accretion flow (\citealt{Sniegowska2020}). 
Alternatively, the changing-look timescale of about one year observed around 2020 could also be attributed to a sudden inflow of surrounding gas 
within the Bondi accretion radius between the BLR and narrow-line region (\citealt{Wang2024}). 
For the first time, we detected a bluer-when-brighter trend in the mid-infrared, which supports the scenario of AGN-heated dust torus (\citealt{Netzer2013}). 
We also noted that the low-ionization broad-line region is very close to the torus, which indicates that BLR may be coupled with the torus. 

\item
During the full-cycle type transition, 
we found a change in the Eddington ratio of Mrk~1018 from 2.0$\times10^{-5}$ to 0.02. 
This suggests a significant state change of the central engine, and implies a transition from a passive to an active state of the SMBH. 
Our findings reveal that the highly variable accretion rate influences the properties of the broad lines.  
For example, the broad-line Balmer decrement varies from about 1.5 to 9, 
initially rising but then declining as the Eddington ratio increases, aligning with the theoretical predictions (\citealt{Netzer1975}). 
The broad Balmer line transitions from a single peak in Type 1.0-1.2 phase to double peaks in Type 1.5-1.8 phase and the separation of the double peak decreases with increasing Eddington ratio, which is observed for the first time. 
Despite significant changes in the central engine over the past 45 years, 
the variation in the broad-line FWHM of Mrk~1018 obeys the relation FWHM$\propto (L_{\rm bol}/{L_{\rm Edd}})^{-0.27}$, 
indicating that the BLR remains virialized and its global kinematics is stable (at least for the low-accretion scenario of Mrk 1018). 
However, the velocity dispersion $\sigma_{\rm line}$ of the broad Balmer lines shows a similar trend as the FWHM when the Eddington ratio is below 
$\sim$0.013 (in Type 1.8-1.5 phase), but sharply increases as Eddington ratio exceeds $\sim$0.013 (in Type 1.2-1.0 phase). 
Consequently, the ratio FWHM/$\sigma_{\rm line}$ remains constant at an average value of 2.96 when the Eddington ratio is below $\sim$0.013, and then drops to 1.71 when the Eddington ratio exceeds this threshold. 
This finding suggests that the state transition of the accretion disk or the rising accretion rate might enhance the turbulent motions in the virialized BLR. 

\item
We estimated the virial mass of the SMBH in Mrk~1018 using the broad H$\beta$ line width (FWHM) and its time delay during the outburst phase, 
yielding $M_{\bullet} = (1.64\pm0.98)\times10^{8}~M_{\sun}$. This is well consistent with the estimate from the $M_{\bullet}-\sigma_{*}$ relation $M_{\bullet} = (1.61\pm0.41)\times 10^{8}~M_{\sun}$.
\end{enumerate}

Understanding the full-cycle changing-look transition and type transition is crucial for studying the feeding and evolution of AGNs. 
Mrk~1018 provides an ideal case for examining the physical processes involved in AGN evolution. 
We expect that the new observations, such as those from the Einstein Probe and the Legacy Survey of Space and Time, 
will unveil more extreme events similar to those observed in Mrk 1018, enhancing our understanding of the nuclear activities. 

\section*{Acknowledgments}
We thank the referee for the useful report that improved the manuscript. 
This work is supported by the National Key R$\&$D Program of China with No. 2021YFA1600404. 
K.X.L. acknowledges financial support from the National Natural Science Foundation of China (NSFC-12073068, 11991051, 11873048, 11703077), 
the Young Talent Project of Yunnan Province, the Youth Innovation Promotion Association of Chinese Academy of Sciences (2022058), 
the Yunnan Province Foundation (202001AT070069), and the Light of West China Program provided by Chinese Academy of Sciences (Y7XB016001), 
the science research grants from the China Manned Space Project with No. CMS-CSST-2021-A06. 
Y.R.L. acknowledges financial support from the NSFC through grant
No. 12273041 and from the Youth Innovation Promotion Association CAS. 
LCH was supported by the National Science Foundation of China (11991052, 12233001), the National Key R\&D Program of China (2022YFF0503401), 
and the China Manned Space Project (CMS-CSST-2021-A04, CMS-CSST-2021-A06). 
H.C.F. acknowledges support from National Natural Science Foundation of China (NSFC-12203096 and 12373018), 
Yunnan Fundamental Research Projects (grant NO. 202301AT070339), 
and Special Research Assistant Funding Project of Chinese Academy of Sciences. 
S.S.L. acknowledges support from National Natural Science Foundation of China (NSFC-12303022), 
Yunnan Fundamental Research Projects (grant No. 202301AT070358), and Yunnan Postdoctoral Research Foundation Funding Project. 
M.Y.S. acknowledges support from the National Natural Science Foundation of China (NSFC-12322303), 
and the Natural Science Foundation of Fujian Province of China (No. 2022J06002). 
C.C. acknowledges NSFC grant No. 12173045. This work is sponsored (in part) by the Chinese Academy of Sciences (CAS), through a grant to the CAS South America Center for Astronomy (CASSACA).

The long-term spectroscopic monitoring of Mrk~1018 has undergone through several years. We acknowledge the support of the staff of the Lijiang 2.4~m telescope. 
Funding for the telescope has been provided by Chinese Academy of Sciences and the People’s Government of Yunnan Province. 

This research makes use of data from Zwicky Transient Facility (ZTF). 
ZTF is supported by the National Science Foundation under grants no. AST-1440341 and AST-2034437 
and a collaboration including current partners Caltech, IPAC, the Weizmann Institute for Science, 
the Oskar Klein Center at Stockholm University, the University of Maryland, 
Deutsches Elektronen-Synchrotron and Humboldt University, the TANGO Con- sortium of Taiwan, 
the University of Wisconsin at Milwaukee, Trinity College Dublin, Lawrence Livermore National Laboratories, IN2P3, 
University of Warwick, Ruhr University Bochum, Northwestern University and former partners the University of Washington, 
Los Alamos National Laboratories, and Lawrence Berkeley National Laboratories. Operations are conducted by COO, IPAC, and UW. 

This research makes use of data from the Asteroid Terrestrial-impact Last Alert System (ATLAS) project. 
The ATLAS project is primarily funded to search for near earth asteroids through NASA grants NN12AR55G, 
80NSSC18K0284, and 80NSSC18K1575; byproducts of the NEO search include images and catalogues from the survey area. 
 
This research makes use of data products from NEOWISE. 
NEOWISE is a project of the Jet Propulsion Laboratory/California Institute of Technology, 
funded by the Planetary Science Division of the National Aeronautics and Space Administration. 

\appendix

\begin{figure*}[htb]
\centering
\includegraphics[angle=0,width=0.85\textwidth]{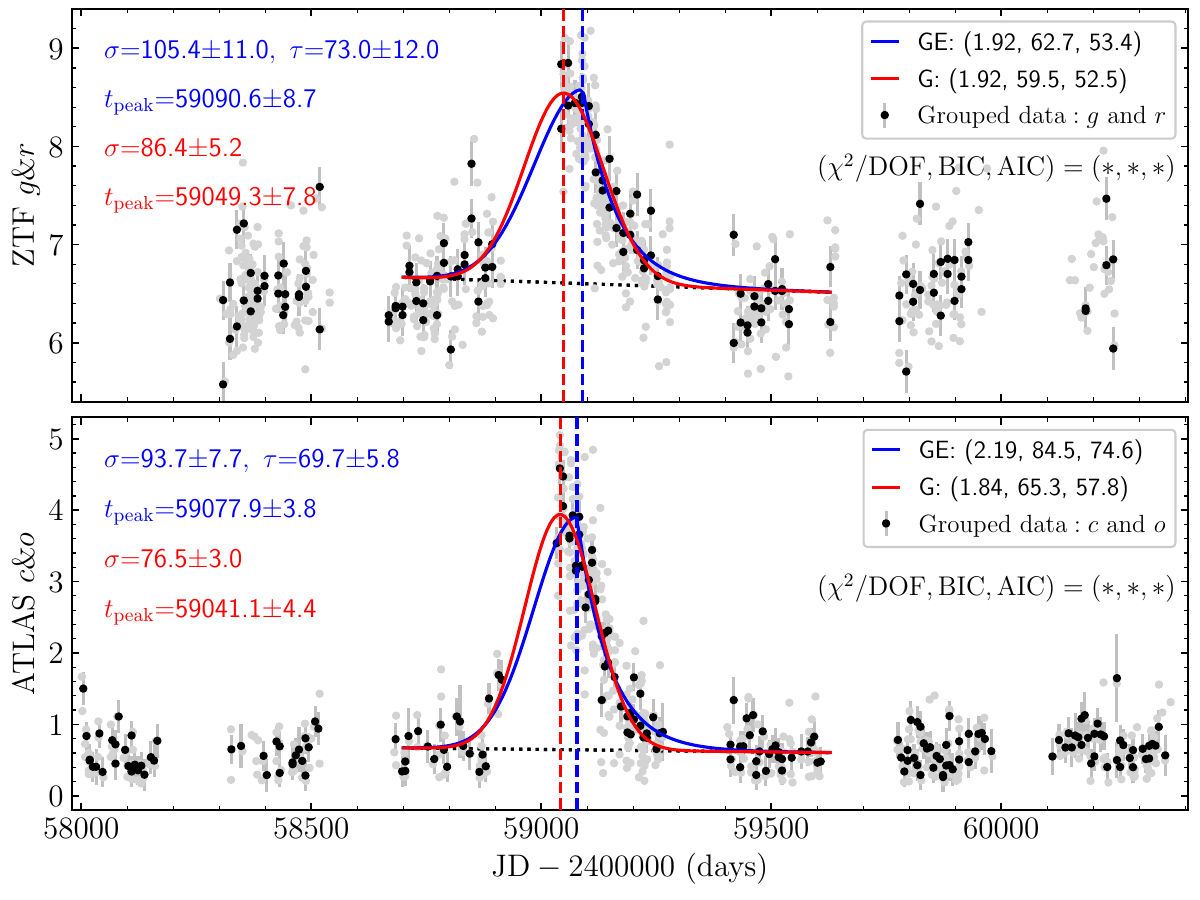}
\caption{
The light curves (in flux density) from ZTF and ATLAS along with model fitting. 
After subtracting the underlying trend, the outburst light-curve profiles are fitted using two models, one with Gaussian rising and exponential dimming (GE; blue lines), and the other with a single Gaussian (G; red lines). 
The best-fit parameters, including $\sigma$ (days), $\tau$ (days) and $t_{\rm peak}$ (-2,400,000~days) are shown in the same color code. 
The vertical dashed line shows the peak position, which has the same color code with model. 
The fitting goodness statistics, including $\chi^{2}$/dof, Bayesian Information Criterion (BIC), and 
Akaike's Information Criterion (AIC) are also noted. 
The Gaussian model yields smaller BIC and AIC values compared to the GE model. 
}
\label{figmodeltest}
\end{figure*}

\section{Outburst light-curve modeling and characteristics} \label{model}
Usually, tidal disruption events (TDEs) and significant changes in accretion rate often drive extreme outburst behaviors 
in galactic centers (\citealt{Graham2020,Zhang2023}). 
The power-law function is widely used to describe the dimming phase of TDEs and generally provides good fits in most cases. 
However, a few TDEs obviously deviate from the power-law decay and need an exponential function to fit the light curves (\citealt{Yao2023}). 
The origins of extreme accretion-driven AGN outburst over various timescales remain contentious and lack a definitive physical model. 
Although supernova events have been observed in some Seyfert host galaxies, there is no established precedent for such occurrences in AGN nuclei, 
so the supernova model is not considered here. 
It should be noted that, \cite{Brogan2023} reported this outburst event, mainly based on the ATLAS photometric data. 
They used a linear function to fit the data in the declining phase, yielding $\chi^2/{\rm dof}>$60, 
and also tried with a parabolic function and a power law with various power-law indexes 
(including -5/3 for a TDE outburst), all yielding $\chi^2/{\rm dof}>$100.
They therefore claimed that the linear function is the best fit. 
By inspecting the multi-band outburst profiles, we find that the dimming phase obviously diverges from the linear trend. 
This prompts us to explore alternative models. We start with Gaussian rising and power-law dimming (GP model): 
\begin{equation}
F_{\lambda}= A\times \left\{
\begin{aligned}
&e^{-(t-t_{\rm peak})/2\sigma^{2}}, t \le t_{\rm peak} \\
&\left(\frac{t-t_{\rm peak}+t_{0}}{t_{0}}\right)^{p}, t> t_{\rm peak}
\end{aligned}
\right.
\end{equation}
where $A$ represents the amplitude, $t$ denotes the observing time, $t_{\rm peak}$ is the light peak time, 
$t_{0}$ is the power-law normalization, and $p$ is the power-law index. 
However, we find that GP model fails to match the data. 
We then attempt to use Gaussian rising plus  exponential dimming (GE model): 
\begin{equation}
F_{\lambda}= A\times \left\{
\begin{aligned}
&e^{-(t-t_{\rm peak})/2\sigma^{2}}, t\le t_{\rm peak} \\
&e^{-(t-t_{\rm peak})/\tau}, t>t_{\rm peak}
\end{aligned}
\right.
\end{equation}
where $\tau$ represents dimming timescale. 
Finally, we also try with a single Gaussian (G model) considering the evident symmetry of the mid-infrared outburst light-curve profile: 
\begin{equation}
F_{\lambda}= A\times e^{-(t-t_{\rm peak})/2\sigma^{2}}, t_{\rm peak}<t\le t_{\rm peak}
\end{equation}

Figure~\ref{figmodeltest} presents the modeling results for merged ZTF and ATLAS light curves. 
The parameters for the GE and Gaussian model (except the amplitudes) are reported, along with the resulting $\chi^{2}$/dof, 
Bayesian Information Criterion (BIC), and Akaike’s Information Criterion (AIC). 
We find that both models yield similar values of $\chi^{2}$/DOF, 
whereas, the BIC and AIC of GE model are slightly larger than Gaussian model. 
Given the nearly symmetric outburst light-curve shapes in the mid-infrared ($W1$ and $W2$ bands), 
we prefer to using the Gaussian model to describe the outburst light-curve profiles. 
Further arguments for this choice are provided in Appendix~\ref{color} and Section~\ref{rob}. 
Consequently, we present the Gaussian modeling results for the outburst light-curve profiles, 
including ZTF optical bands, broad Balmer lines, and mid-infrared bands in Figure~\ref{cle1}, 
from which we derive the outburst characteristics listed in Table~\ref{tab2}. 

\begin{figure*}[htb]
\centering
\includegraphics[angle=0,width=0.48\textwidth]{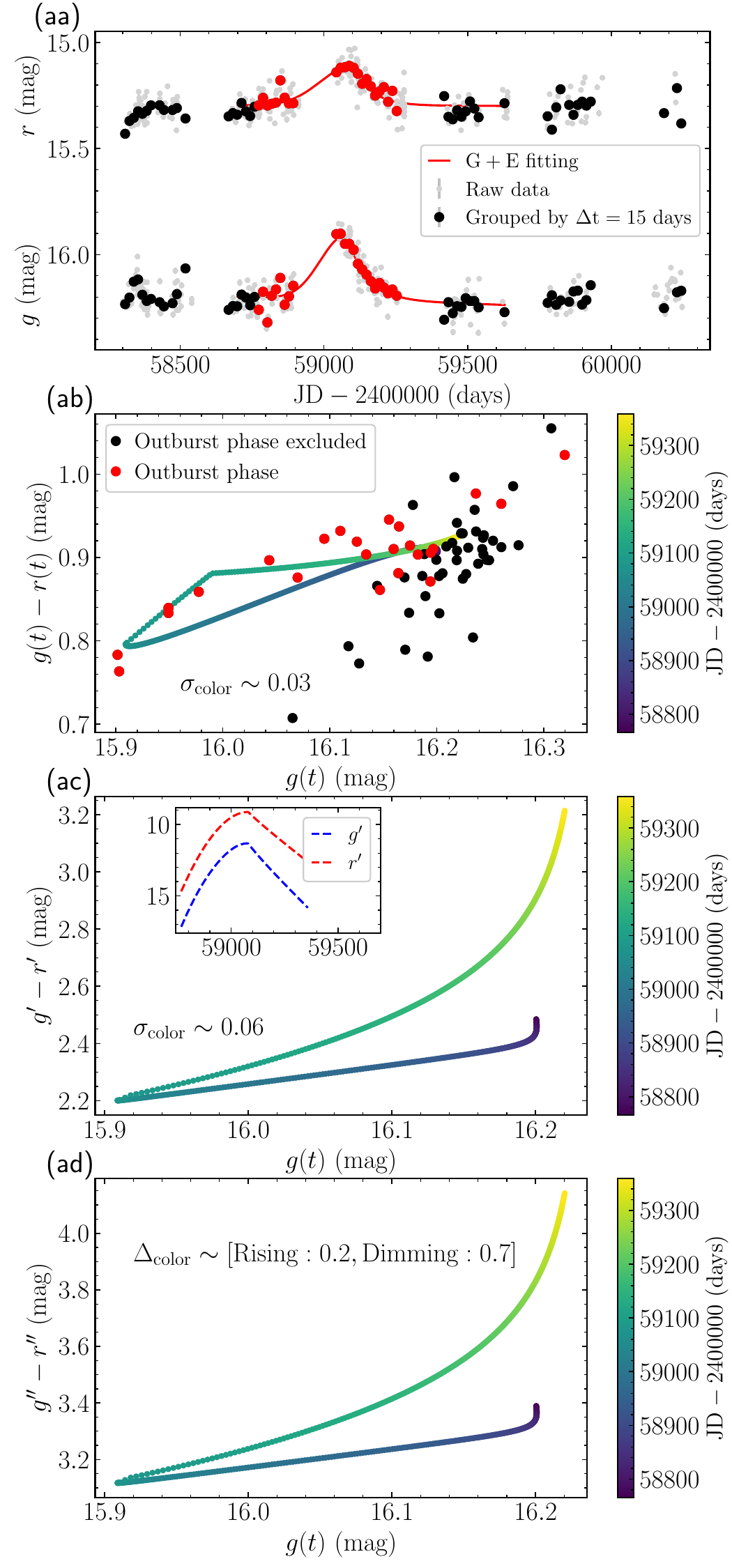}
\includegraphics[angle=0,width=0.48\textwidth]{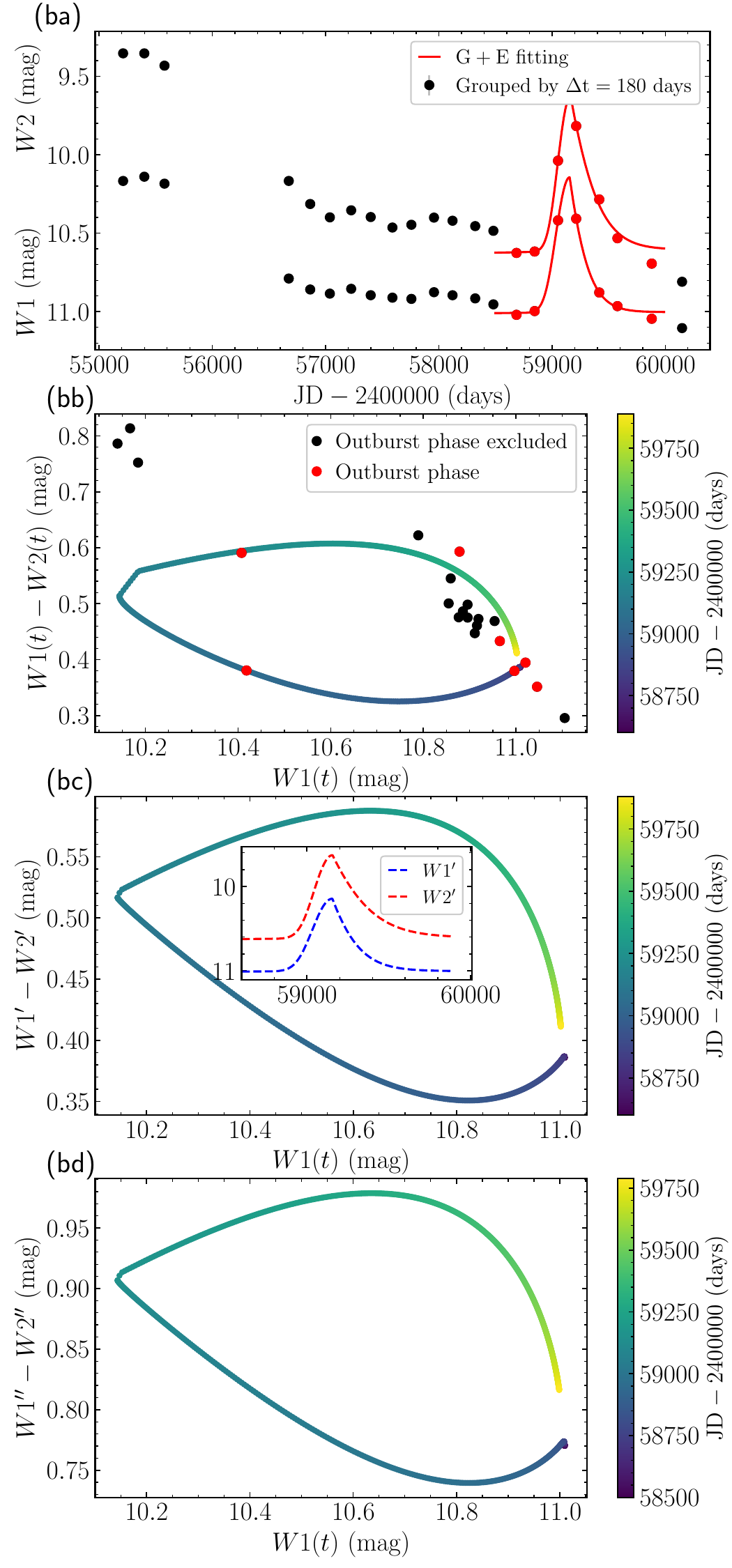}
\caption{
Same as Figure~\ref{figcolor}, but for the optical and mid-infrared color relations based on GE model (also see Appendix~\ref{color}). 
}
\label{figsp2}
\end{figure*}

\section{Color vs. magnitude diagrams} \label{color} 
To analyze the color$-$magnitude relations in the optical and mid-infrared bands, we grouped the raw $g$- and $r$-band photometric data by a 15-day intervals and the raw $W1$- and $W2$-band
photometric data by a  six-month interval. 
Appendix~\ref{model} (also see Figure~\ref{figmodeltest}) suggests that the outburst light-curve profiles 
can be effectively modeled using either the Gaussian or GE model, 
with the Gaussian model yielding slightly lower BIC and AIC values. 
In this appendix, we examine the color$-$magnitude relations during the outburst phase from both models. 

The results with the Gaussian models are shown in the top panels of Figure~\ref{figcolor} 
(panels aa and ba; also see Appendix~\ref{model} and Figure~\ref{cle1}). 
The second row (panels ab and bb) of Figure~\ref{figcolor} presents the relations of the color $g$-$r$ versus $g$ and $W1$-$W2$ versus $W1$, 
which are generated directly from the grouped and fitting model. 
The intrinsic optical color is contaminated by delayed emission lines because for Mrk~1018 ($z$=0.043), the ZTF $g$-band covers the broad H$\beta$ lines 
and ZTF $r$-band covers the broad H$\alpha$ lines. 
Additionally, the intrinsic mid-infrared color is affected by inter-band time delay. 
The intrinsic optical color ($g{'}$-$r{'}$) is derived by adjusting $g(t)$-$r(t)$ with a correction factor of 
-2.5log$\frac{{\rm H\alpha}(t-\tau_{\rm H\alpha})}{{\rm H\beta}(t-\tau_{\rm H\beta})}$, 
thus $g{'}$-$r{'}$=$g(t)$-$r(t-\tau_{r})$-(-2.5log$\frac{{\rm H\alpha}(t-\tau_{\rm H\alpha})}{{\rm H\beta}(t-\tau_{\rm H\beta})})$, 
where $\tau_{r}$, $\tau_{\rm H\alpha}$ and $\tau_{\rm H\beta}$ are the time delays of the ZTF $r$-band, broad H$\alpha$ and H$\beta$ lines 
with respect to the varying ZTF $g$-band.  
The intrinsic mid-infrared color ($W1{'}$-$W2{'}$) is obtained by correcting the inter-band time delay ($\tau_{W2}$) between the $W2$-band 
and the varying $W1$-band, given by $W1{'}$-$W2{'}$=$W1(t)$-$W2(t-\tau_{W2})$. 
In the third row (panels ac and bc) of Figure~\ref{figcolor}, the colorized thick lines and color bars present the intrinsic color$-$magnitude relations 
for optical and mid-infrared bands during the outburst phase. 
In order to obtain the net color$-$magnitude relation that reflects the nature of the outburst phase, 
we removed the underlying linear variation trend from 
the intrinsic color$-$magnitude relations given above. 
The resulting net optical color$-$magnitude relation is defined as 
$g{''}$-$r{''}$ = $g{'}$-$r{'}$-(-2.5log$\frac{\delta r(t-\tau_{r})}{\delta g(t)}$), 
where $\delta r$ and $\delta g$ represent the linear variation trends beneath the outburst light-curve profiles for the $r$- and $g$-bands, 
assuming equal zero-point flux densities for both bands.  
The net mid-infrared color$-$magnitude relation is expressed as $W1{''}$-$W1{''}$ = $W1{'}$-$W2{'}$-(-2.5log$\frac{\delta 2(t-\tau_{w2})/2.43}{\delta 1(t)/8.03}$), 
where $\delta 2$ and $\delta 1$ represent the linear variation trends beneath the outburst light-curve profiles for $W2$- and $W1$-bands, 
and 2.43/8.03 is the zero-point flux density ratio. 
We present net optical and mid-infrared color$-$magnitude relations in the bottommost row (panels ad and bd). 
The corresponding results from the GE model are shown in Figure~\ref{figsp2}. 

We find that in the Gaussian model, the intrinsic color$-$magnitude relations 
during rising and dimming phases show nearly identical slopes and intercepts in both optical and mid-infrared bands. 
In contrast, in the GE model,
the intrinsic color$-$magnitude relations in optical band show different amplitudes and slopes between the rising and dimming phases, 
while the mid-infrared relations shows opposite evolutionary tracks. 
Nonetheless, during the dimming phase, both models give consistent color$-$magnitude relations, 
with optical colors becoming redder as the source dims, suggesting these trends are model-independent.  
Additionally, the optical color$-$magnitude relation in the dimming phase does not not coincide with theoretical prediction of TDEs, 
which typically exhibit a bluer spectrum during dimming, thereby excluding the TDE scenario in Mrk 1018. 



\begin{thebibliography}{99}
\bibitem[Abazajian et al.(2009)]{Abazajian2009} Abazajian, K.~N., Adelman-McCarthy, J.~K., Ag{\"u}eros, M.~A., et al.\ 2009, \apjs, 182, 543. doi:10.1088/0067-0049/182/2/543

\bibitem[Abramowicz et al.(1995)]{Abramowicz1995} Abramowicz, M.~A., Chen, X., Kato, S., et al.\ 1995, \apjl, 438, L37. doi:10.1086/187709

\bibitem[Antonucci(1993)]{Antonucci1993} Antonucci, R.\ 1993, \araa, 31, 473. doi:10.1146/annurev.aa.31.090193.002353

\bibitem[Barth et al.(2015)]{Barth2015} Barth, A.~J., Bennert, V.~N., Canalizo, G., et al.\ 2015, \apjs, 217, 26. doi:10.1088/0067-0049/217/2/26

\bibitem[Bennert et al.(2011)]{Bennert2011} Bennert, V.~N., Auger, M.~W., Treu, T., et al.\ 2011, \apj, 742, 107. doi:10.1088/0004-637X/742/2/107 

\bibitem[Bentz et al.(2013)]{Bentz2013} Bentz, M.~C., Denney, K.~D., Grier, C.~J., et al.\ 2013, \apj, 767, 149. doi:10.1088/0004-637X/767/2/149

\bibitem[Bentz et al.(2010)]{Bentz2010} Bentz, M.~C., Walsh, J.~L., Barth, A.~J., et al.\ 2010, \apj, 716, 993. doi:10.1088/0004-637X/716/2/993

\bibitem[Blandford \& McKee(1982)]{Blandford1982} Blandford, R.~D. \& McKee, C.~F.\ 1982, \apj, 255, 419. doi:10.1086/159843

\bibitem[Boroson \& Green(1992)]{Boroson1992} Boroson, T.~A. \& Green, R.~F.\ 1992, \apjs, 80, 109. doi:10.1086/191661

\bibitem[Brogan et al.(2023)]{Brogan2023} Brogan, R., Krumpe, M., Homan, D., et al.\ 2023, \aap, 677, A116. doi:10.1051/0004-6361/202346475

\bibitem[Bruzual \& Charlot(2003)]{Bruzual2003} Bruzual, G. \& Charlot, S.\ 2003, \mnras, 344, 1000. doi:10.1046/j.1365-8711.2003.06897.x

\bibitem[Cai et al.(2016)]{Cai2016} Cai, Z.-Y., Wang, J.-X., Gu, W.-M., et al.\ 2016, \apj, 826, 7. doi:10.3847/0004-637X/826/1/7

\bibitem[Capellupo et al.(2015)]{Capellupo2015} Capellupo, D.~M., Netzer, H., Lira, P., et al.\ 2015, \mnras, 446, 3427. doi:10.1093/mnras/stu2266

\bibitem[Chen \& Halpern(1989)]{Chen1989} Chen, K. \& Halpern, J.~P.\ 1989, \apj, 344, 115. doi:10.1086/167782

\bibitem[Chen et al.(2023a)]{Chen2023a} Chen, Y.-J., Bao, D.-W., Zhai, S., et al.\ 2023a, \mnras, 520, 1807. doi:10.1093/mnras/stad051

\bibitem[Chen et al.(2023b)]{Chen2023b} Chen, Y.-J., Liu, J.-R., Zhai, S., et al.\ 2023b, \mnras, 522, 3439. doi:10.1093/mnras/stad1136

\bibitem[Cho et al.(2023)]{Cho2023} Cho, H., Woo, J.-H., Wang, S., et al.\ 2023, \apj, 953, 142. doi:10.3847/1538-4357/ace1e5

\bibitem[Cohen et al.(1986)]{Cohen1986} Cohen, R.~D., Rudy, R.~J., Puetter, R.~C., et al.\ 1986, \apj, 311, 135. doi:10.1086/164758

\bibitem[Cutri et al.(1985)]{Cutri1985} Cutri, R.~M., Wisniewski, W.~Z., Rieke, G.~H., et al.\ 1985, \apj, 296, 423. doi:10.1086/163461

\bibitem[Czerny \& Hryniewicz(2011)]{Czerny2011} Czerny, B. \& Hryniewicz, K.\ 2011, \aap, 525, L8. doi:10.1051/0004-6361/201016025

\bibitem[Czerny et al.(2023)]{Czerny2023} Czerny, B., Panda, S., Prince, R., et al.\ 2023, \aap, 675, A163. doi:10.1051/0004-6361/202345844

\bibitem[Denney et al.(2014)]{Denney2014} Denney, K.~D., De Rosa, G., Croxall, K., et al.\ 2014, \apj, 796, 134. doi:10.1088/0004-637X/796/2/134

\bibitem[Dong et al.(2011)]{Dong2011} Dong, X.-B., Wang, J.-G., Ho, L.~C., et al.\ 2011, \apj, 736, 86. doi:10.1088/0004-637X/736/2/86

\bibitem[Du et al.(2016)]{Du2016} Du, P., Lu, K.-X., Zhang, Z.-X., et al.\ 2016, \apj, 825, 126. doi:10.3847/0004-637X/825/2/126

\bibitem[Du et al.(2018)]{Du2018} Du, P., Zhang, Z.-X., Wang, K., et al.\ 2018, \apj, 856, 6. doi:10.3847/1538-4357/aaae6b

\bibitem[Elitzur et al.(2014)]{Elitzur2014} Elitzur, M., Ho, L.~C., \& Trump, J.~R.\ 2014, \mnras, 438, 3340. doi:10.1093/mnras/stt2445 

\bibitem[Fan et al.(2015)]{Fan2015} Fan, Y.-F., Bai, J.-M., Zhang, J.-J., et al.\ 2015, Research in Astronomy and Astrophysics, 15, 918. doi:10.1088/1674-4527/15/6/014

\bibitem[Feng et al.(2024)]{Feng2024} Feng, H.-C., Li, S.-S., Bai, J.~M., et al.\ 2024, arXiv:2409.01637. doi:10.48550/arXiv.2409.01637

\bibitem[Feng et al.(2021)]{Feng2021} Feng, H.-C., Liu, H.~T., Bai, J.~M., et al.\ 2021, \apj, 912, 92. doi:10.3847/1538-4357/abefe0

\bibitem[Ferrarese \& Merritt(2000)]{Ferrarese2000} Ferrarese, L. \& Merritt, D.\ 2000, \apjl, 539, L9. doi:10.1086/312838

\bibitem[Gaspari et al.(2020)]{Gaspari2020} Gaspari, M., Tombesi, F., \& Cappi, M.\ 2020, Nature Astronomy, 4, 10. doi:10.1038/s41550-019-0970-1

\bibitem[Gebhardt et al.(2000)]{Gebhardt2000} Gebhardt, K., Bender, R., Bower, G., et al.\ 2000, \apjl, 539, L13. doi:10.1086/312840

\bibitem[Goad et al.(2012)]{Goad2012} Goad, M.~R., Korista, K.~T., \& Ruff, A.~J.\ 2012, \mnras, 426, 3086. doi:10.1111/j.1365-2966.2012.21808.x

\bibitem[Graham et al.(2019)]{Graham2019} Graham, M.~J., Kulkarni, S.~R., Bellm, E.~C., et al.\ 2019, \pasp, 131, 078001. doi:10.1088/1538-3873/ab006c

\bibitem[Graham et al.(2020)]{Graham2020} Graham, M.~J., Ross, N.~P., Stern, D., et al.\ 2020, \mnras, 491, 4925. doi:10.1093/mnras/stz3244

\bibitem[Greene \& Ho(2005)]{Greene2005} Greene, J.~E. \& Ho, L.~C.\ 2005, \apj, 630, 122. doi:10.1086/431897

\bibitem[Greene et al.(2020)]{Greene2020} Greene, J.~E., Strader, J., \& Ho, L.~C.\ 2020, \araa, 58, 257. doi:10.1146/annurev-astro-032620-021835

\bibitem[Grier et al.(2017)]{Grier2017} Grier, C.~J., Trump, J.~R., Shen, Y., et al.\ 2017, \apj, 851, 21. doi:10.3847/1538-4357/aa98dc

\bibitem[Guo \& Gu(2014)]{Guo2014} Guo, H. \& Gu, M.\ 2014, \apj, 792, 33. doi:10.1088/0004-637X/792/1/33

\bibitem[Guo \& Gu(2016)]{Guo2016} Guo, H. \& Gu, M.\ 2016, \apj, 822, 26. doi:10.3847/0004-637X/822/1/26

\bibitem[Guo et al.(2024a)]{Guo2024a} Guo, W.-J., Zou, H., Fawcett, V.~A., et al.\ 2024a, \apjs, 270, 26. doi:10.3847/1538-4365/ad118a
\bibitem[Guo et al.(2024b)]{Guo2024b} Guo, W.-J., Zou, H., Greenwell, C.~L., et al.\ 2024b, arXiv:2408.00402. doi:10.48550/arXiv.2408.00402

\bibitem[Ho(2008)]{Ho2008} Ho, L.~C.\ 2008, \araa, 46, 475. doi:10.1146/annurev.astro.45.051806.110546

\bibitem[Ho(2009)]{Ho2009} Ho, L.~C.\ 2009, \apj, 699, 638. doi:10.1088/0004-637X/699/1/638

\bibitem[Ho et al.(2000)]{Ho2000} Ho, L.~C., Rudnick, G., Rix, H.-W., et al.\ 2000, \apj, 541, 120. doi:10.1086/309440

\bibitem[Holoien et al.(2016)]{Holoien2016} Holoien, T.~W.-S., Kochanek, C.~S., Prieto, J.~L., et al.\ 2016, \mnras, 455, 2918. doi:10.1093/mnras/stv2486

\bibitem[Hu et al.(2015)]{Hu2015} Hu, C., Du, P., Lu, K.-X., et al.\ 2015, \apj, 804, 138. doi:10.1088/0004-637X/804/2/138

\bibitem[Hu et al.(2020)]{Hu2020} Hu, C., Li, S.-S., Guo, W.-J., et al.\ 2020, \apj, 905, 75. doi:10.3847/1538-4357/abc2da

\bibitem[Husemann et al.(2016)]{Husemann2016} Husemann, B., Urrutia, T., Tremblay, G.~R., et al.\ 2016, \aap, 593, L9. doi:10.1051/0004-6361/201629245

\bibitem[Husser et al.(2013)]{Husser2013} Husser, T.-O., Wende-von Berg, S., Dreizler, S., et al.\ 2013, \aap, 553, A6. doi:10.1051/0004-6361/201219058

\bibitem[Jarrett et al.(2013)]{Jarrett2013} Jarrett, T.~H., Masci, F., Tsai, C.~W., et al.\ 2013, \aj, 145, 6. doi:10.1088/0004-6256/145/1/6

\bibitem[Jones et al.(2009)]{Jones2009} Jones, D.~H., Read, M.~A., Saunders, W., et al.\ 2009, \mnras, 399, 683. doi:10.1111/j.1365-2966.2009.15338.x

\bibitem[Kewley et al.(2001)]{Kewley2001} Kewley, L.~J., Dopita, M.~A., Sutherland, R.~S., et al.\ 2001, \apj, 556, 121. doi:10.1086/321545

\bibitem[Khachikian \& Weedman(1971)]{Khachikian1971} Khachikian, E.~Y. \& Weedman, D.~W.\ 1971, \apjl, 164, L109. doi:10.1086/180701

\bibitem[Kim et al.(2018)]{Kim2018} Kim, D.-C., Yoon, I., \& Evans, A.~S.\ 2018, \apj, 861, 51. doi:10.3847/1538-4357/aac77d

\bibitem[Kollatschny \& Zetzl(2011)]{Kollatschny2011} Kollatschny, W. \& Zetzl, M.\ 2011, \nat, 470, 366. doi:10.1038/nature09761

\bibitem[Kollatschny \& Zetzl(2013)]{Kollatschny2013} Kollatschny, W. \& Zetzl, M.\ 2013, \aap, 549, A100. doi:10.1051/0004-6361/201219411

\bibitem[Komossa et al.(2024)]{Komossa2024} Komossa, S., Grupe, D., Marziani, P., et al.\ 2024, arXiv:2408.00089. doi:10.48550/arXiv.2408.00089

\bibitem[Kong \& Ho(2018)]{Kong2018} Kong, M. \& Ho, L.~C.\ 2018, \apj, 859, 116. doi:10.3847/1538-4357/aabe2a

\bibitem[Korista \& Goad(2004)]{Korista2004} Korista, K.~T. \& Goad, M.~R.\ 2004, \apj, 606, 749. doi:10.1086/383193

\bibitem[Koss et al.(2011)]{Koss2011} Koss, M., Mushotzky, R., Veilleux, S., et al.\ 2011, \apj, 739, 57. doi:10.1088/0004-637X/739/2/57

\bibitem[Kova{\v{c}}evi{\'c} et al.(2010)]{Kova2010} Kova{\v{c}}evi{\'c}, J., Popovi{\'c}, L. {\v{C}}., \& Dimitrijevi{\'c}, M.~S.\ 2010, \apjs, 189, 15

\bibitem[Krumpe et al.(2017)]{Krumpe2017} Krumpe, M., Husemann, B., Tremblay, G.~R., et al.\ 2017, \aap, 607, L9. doi:10.1051/0004-6361/201731967

\bibitem[Laha et al.(2022)]{Laha2022} Laha, S., Meyer, E., Roychowdhury, A., et al.\ 2022, \apj, 931, 5. doi:10.3847/1538-4357/ac63aa

\bibitem[Leloudas et al.(2019)]{Leloudas2019} Leloudas, G., Dai, L., Arcavi, I., et al.\ 2019, \apj, 887, 218. doi:10.3847/1538-4357/ab5792

\bibitem[Li et al.(2022)]{Li2022} Li, R., Ho, L.~C., Ricci, C., et al.\ 2022, \apj, 933, 70. doi:10.3847/1538-4357/ac714a

\bibitem[Li et al.(2024)]{Li2024} Li, S.-S., Feng, H.-C., Liu, H.~T., et al.\ 2024, arXiv:2407.05414. doi:10.48550/arXiv.2407.05414

\bibitem[Liu et al.(2021)]{Liu2021} Liu, W.-J., Lira, P., Yao, S., et al.\ 2021, \apj, 915, 63. doi:10.3847/1538-4357/abf82c

\bibitem[Li et al.(2014)]{LYR2014} Li, Y.-R., Wang, J.-M., Hu, C., et al.\ 2014, \apjl, 786, L6. doi:10.1088/2041-8205/786/1/L6

\bibitem[Lu et al.(2022)]{Lu2022} Lu, K.-X., Bai, J.-M., Wang, J.-M., et al.\ 2022, \apjs, 263, 10. doi:10.3847/1538-4365/ac94d3 

\bibitem[Lu et al.(2019a)]{Lu2019a} Lu, K.-X., Bai, J.-M., Zhang, Z.-X., et al.\ 2019a, \apj, 887, 135. doi:10.3847/1538-4357/ab5790

\bibitem[Lu et al.(2021a)]{Lu2021a} Lu, K.-X., Wang, J.-G., Zhang, Z.-X., et al.\ 2021a, \apj, 918, 50. doi:10.3847/1538-4357/ac0c78

\bibitem[Lu et al.(2019c)]{Lu2019b} Lu, K.-X., Zhao, Y., Bai, J.-M., et al.\ 2019c, \mnras, 483, 1722 . doi:10.1093/mnras/sty3229

\bibitem[Lu et al.(2021b)]{Lu2021b} Lu, K.-X., Zhang, Z.-X., Huang, Y.-K., et al.\ 2021b, Research in Astronomy and Astrophysics, 21, 183. doi:10.1088/1674-4527/21/7/183

\bibitem[Lyu et al.(2021)]{Lyu2021} Lyu, B., Yan, Z., Yu, W., et al.\ 2021, \mnras, 506, 4188. doi:10.1093/mnras/stab1581

\bibitem[MacLeod et al.(2016)]{MacLeod2016} MacLeod, C.~L., Ross, N.~P., Lawrence, A., et al.\ 2016, \mnras, 457, 389. doi:10.1093/mnras/stv2997

\bibitem[Markwardt(2009)]{Markwardt2009} Markwardt, C.~B.\ 2009, Astronomical Data Analysis Software and Systems XVIII, 411, 251. doi:10.48550/arXiv.0902.2850

\bibitem[Matt et al.(2003)]{Matt2003} Matt, G., Guainazzi, M., \& Maiolino, R.\ 2003, \mnras, 342, 422. doi:10.1046/j.1365-8711.2003.06539.x

\bibitem[Ma et al.(2023)]{Ma2023} Ma, Y.-S., Li, S.-J., Gu, C.-S., et al.\ 2023, \mnras, 522, 5680. doi:10.1093/mnras/stad1377

\bibitem[Malik et al.(2024)]{Malik2024} Malik, U., Sharp, R., Penton, A., et al.\ 2024, \mnras, 531, 163. doi:10.1093/mnras/stae1154

\bibitem[McElroy et al.(2016)]{McElroy2016} McElroy, R.~E., Husemann, B., Croom, S.~M., et al.\ 2016, \aap, 593, L8. doi:10.1051/0004-6361/201629102

\bibitem[McGill et al.(2008)]{McGill2008} McGill, K.~L., Woo, J.-H., Treu, T., et al.\ 2008, \apj, 673, 703. doi:10.1086/524349

\bibitem[McLure \& Dunlop(2004)]{McLure2004} McLure, R.~J. \& Dunlop, J.~S.\ 2004, \mnras, 352, 1390. doi:10.1111/j.1365-2966.2004.08034.x

\bibitem[Mereghetti et al.(2021)]{Mereghetti2021} Mereghetti, S., Balman, S., Caballero-Garcia, M., et al.\ 2021, Experimental Astronomy, 52, 309. doi:10.1007/s10686-021-09809-6

\bibitem[Merloni et al.(2015)]{Merloni2015} Merloni, A., Dwelly, T., Salvato, M., et al.\ 2015, \mnras, 452, 69. doi:10.1093/mnras/stv1095

\bibitem[Moran et al.(1996)]{Moran1996} Moran, E.~C., Halpern, J.~P., \& Helfand, D.~J.\ 1996, \apjs, 106, 341. doi:10.1086/192341

\bibitem[Naddaf \& Czerny(2022)]{Naddaf2022} Naddaf, M.~H. \& Czerny, B.\ 2022, \aap, 663, A77. doi:10.1051/0004-6361/202142806

\bibitem[Nagoshi et al.(2024)]{Nagoshi2024} Nagoshi, S., Iwamuro, F., Yamada, S., et al.\ 2024, \mnras, 529, 393. doi:10.1093/mnras/stae319

\bibitem[Nelson \& Whittle(1996)]{Nelson1996} Nelson, C.~H. \& Whittle, M.\ 1996, \apj, 465, 96. doi:10.1086/177405

\bibitem[Netzer(1975)]{Netzer1975} Netzer, H.\ 1975, \mnras, 171, 395. doi:10.1093/mnras/171.2.395

\bibitem[Netzer(2013)]{Netzer2013} Netzer, H.\ 2013, The Physics and Evolution of Active Galactic Nuclei, by Hagai Netzer, Cambridge, UK: Cambridge University Press, 2013

\bibitem[Netzer(2015)]{Netzer2015} Netzer, H.\ 2015, \araa, 53, 365. doi:10.1146/annurev-astro-082214-122302

\bibitem[Oknyansky et al.(2019)]{Oknyansky2019} Oknyansky, V.~L., Winkler, H., Tsygankov, S.~S., et al.\ 2019, \mnras, 483, 558. doi:10.1093/mnras/sty3133

\bibitem[Osterbrock(1977)]{Osterbrock1977} Osterbrock, D.~E.\ 1977, \apj, 215, 733. doi:10.1086/155407

\bibitem[Osterbrock(1978)]{Osterbrock1978} Osterbrock, D.~E.\ 1978, Proceedings of the National Academy of Science, 75, 540. doi:10.1073/pnas.75.2.540

\bibitem[Osterbrock(1981)]{Osterbrock1981} Osterbrock, D.~E.\ 1981, \apj, 249, 462. doi:10.1086/159306

\bibitem[Osterbrock \& Ferland(2006)]{Osterbrock2006} Osterbrock, D.~E. \& Ferland, G.~J.\ 2006, Astrophysics of gaseous nebulae and active galactic nuclei, 2nd. ed. by D.E. Osterbrock and G.J. Ferland. Sausalito, CA: University Science Books, 2006

\bibitem[Panda \& {\'S}niegowska(2024)]{Panda2024} Panda, S. \& {\'S}niegowska, M.\ 2024, \apjs, 272, 13. doi:10.3847/1538-4365/ad344f

\bibitem[Panessa \& Bassani(2002)]{Panessa2002} Panessa, F. \& Bassani, L.\ 2002, \aap, 394, 435. doi:10.1051/0004-6361:20021161

\bibitem[Park et al.(2022)]{Park2022} Park, D., Barth, A.~J., Ho, L.~C., et al.\ 2022, \apjs, 258, 38. doi:10.3847/1538-4365/ac3f3e

\bibitem[Park et al.(2012)]{Park2012} Park, D., Woo, J.-H., Treu, T., et al.\ 2012, \apj, 747, 30. doi:10.1088/0004-637X/747/1/30

\bibitem[Penston \& Perez(1984)]{Penston1984} Penston, M.~V. \& Perez, E.\ 1984, \mnras, 211, 33P. doi:10.1093/mnras/211.1.33P

\bibitem[Peterson(1993)]{Peterson1993} Peterson, B.~M.\ 1993, \pasp, 105, 247. doi:10.1086/133140
\bibitem[Peterson et al.(2004)]{Peterson2004} Peterson, B.~M., Ferrarese, L., Gilbert, K.~M., et al.\ 2004, \apj, 613, 682. doi:10.1086/423269

\bibitem[Ren et al.(2022)]{Ren2022} Ren, W., Wang, J., Cai, Z., et al.\ 2022, \apj, 925, 50. doi:10.3847/1538-4357/ac3828

\bibitem[Ricci \& Trakhtenbrot(2023)]{Ricci2023} Ricci, C. \& Trakhtenbrot, B.\ 2023, Nature Astronomy, 7, 1282. doi:10.1038/s41550-023-02108-4

\bibitem[Runnoe et al.(2016)]{Runnoe2016} Runnoe, J.~C., Cales, S., Ruan, J.~J., et al.\ 2016, \mnras, 455, 1691. doi:10.1093/mnras/stv2385

\bibitem[Sakata et al.(2011)]{Sakata 2011} Sakata, Y., Morokuma, T., Minezaki, T., et al.\ 2011, \apj, 731, 50. doi:10.1088/0004-637X/731/1/50

\bibitem[Shakura \& Sunyaev(1973)]{Shakura1973} Shakura, N.~I. \& Sunyaev, R.~A.\ 1973, \aap, 24, 337

\bibitem[Shapovalova et al.(2010)]{Shapovalova2010} Shapovalova, A.~I., Popovi{\'c}, L. {\v{C}}., Burenkov, A.~N., et al.\ 2010, \aap, 509, A106. doi:10.1051/0004-6361/200912311

\bibitem[Shappee et al.(2014)]{Shappee2014} Shappee, B.~J., Prieto, J.~L., Grupe, D., et al.\ 2014, \apj, 788, 48. doi:10.1088/0004-637X/788/1/48

\bibitem[Sheng et al.(2017)]{Sheng2017} Sheng, Z., Wang, T., Jiang, N., et al.\ 2017, \apjl, 846, L7. doi:10.3847/2041-8213/aa85de

\bibitem[Sniegowska et al.(2020)]{Sniegowska2020} Sniegowska, M., Czerny, B., Bon, E., et al.\ 2020, \aap, 641, A167. doi:10.1051/0004-6361/202038575

\bibitem[Stern \& Laor(2012a)]{Stern2012a} Stern, J. \& Laor, A.\ 2012a, \mnras, 423, 600. doi:10.1111/j.1365-2966.2012.20901.x

\bibitem[Stern \& Laor(2012b)]{Stern2012b} Stern, J. \& Laor, A.\ 2012b, \mnras, 426, 2703. doi:10.1111/j.1365-2966.2012.21772.x

\bibitem[Strateva et al.(2003)]{Strateva2003} Strateva, I.~V., Strauss, M.~A., Hao, L., et al.\ 2003, \aj, 126, 1720. doi:10.1086/378367

\bibitem[Sulentic et al.(2002)]{Sulentic2002} Sulentic, J.~W., Marziani, P., Zamanov, R., et al.\ 2002, \apjl, 566, L71. doi:10.1086/339594

\bibitem[Sun et al.(2018)]{Sun2018} Sun, M., Grier, C.~J., \& Peterson, B.~M.\ 2018, Astrophysics Source Code Library. ascl:1805.032

\bibitem[Tonry et al.(2018)]{Tonry2018} Tonry, J.~L., Denneau, L., Heinze, A.~N., et al.\ 2018, \pasp, 130, 064505. doi:10.1088/1538-3873/aabadf

\bibitem[Trakhtenbrot et al.(2019)]{Trakhtenbrot2019} Trakhtenbrot, B., Arcavi, I., MacLeod, C.~L., et al.\ 2019, \apj, 883, 94. doi:10.3847/1538-4357/ab39e4

\bibitem[Tran et al.(2011)]{Tran2011} Tran, H.~D., Lyke, J.~E., \& Mader, J.~A.\ 2011, \apjl, 726, L21. doi:10.1088/2041-8205/726/2/L21

\bibitem[Trippe et al.(2010)]{Trippe2010} Trippe, M.~L., Crenshaw, D.~M., Deo, R.~P., et al.\ 2010, \apj, 725, 1749. doi:10.1088/0004-637X/725/2/1749

\bibitem[Urry \& Padovani(1995)]{Urry1995} Urry, C.~M. \& Padovani, P.\ 1995, \pasp, 107, 803. doi:10.1086/133630

\bibitem[V{\'e}ron-Cetty et al.(2004)]{Veron2004} V{\'e}ron-Cetty, M.-P., Joly, M., \& V{\'e}ron, P.\ 2004, \aap, 417, 515

\bibitem[V{\'e}ron-Cetty et al.(2001)]{Veron2001} V{\'e}ron-Cetty, M.-P., V{\'e}ron, P., \& Gon{\c{c}}alves, A.~C.\ 2001, \aap, 372, 730. doi:10.1051/0004-6361:20010489

\bibitem[Veronese et al.(2024)]{Veronese2024} Veronese, S., Vignali, C., Severgnini, P., et al.\ 2024, \aap, 683, A131. doi:10.1051/0004-6361/202348098

\bibitem[Wang et al.(2019)]{Wang2019} Wang, C.-J., Bai, J.-M., Fan, Y.-F., et al.\ 2019, Research in Astronomy and Astrophysics, 19, 149. doi:10.1088/1674-4527/19/10/149

\bibitem[Wang et al.(2017)]{Wang2017} Wang, J.-M., Du, P., Brotherton, M.~S., et al.\ 2017, Nature Astronomy, 1, 775. doi:10.1038/s41550-017-0264-4

\bibitem[Wang et al.(2024)]{Wang2024} Wang, J., Xu, D.~W., Cao, X., et al.\ 2024, \apj, 970, 85. doi:10.3847/1538-4357/ad4d89 

\bibitem[Winkler(1992)]{Winkler1992} Winkler, H.\ 1992, \mnras, 257, 677. doi:10.1093/mnras/257.4.677

\bibitem[Wright et al.(2010)]{Wright2010} Wright, E.~L., Eisenhardt, P.~R.~M., Mainzer, A.~K., et al.\ 2010, \aj, 140, 1868. doi:10.1088/0004-6256/140/6/1868

\bibitem[Wu et al.(2023)]{Wu2023} Wu, J., Wu, Q., Xue, H., et al.\ 2023, \apj, 950, 106. doi:10.3847/1538-4357/acce9e

\bibitem[Xu et al.(2024)]{Xu2024} Xu, D.~W., Komossa, S., Grupe, D., et al.\ 2024, Universe, 10, 61. doi:10.3390/universe10020061

\bibitem[Yang et al.(2018)]{Yang2018} Yang, Q., Wu, X.-B., Fan, X., et al.\ 2018, \apj, 862, 109. doi:10.3847/1538-4357/aaca3a

\bibitem[Yao et al.(2023)]{Yao2023} Yao, Y., Ravi, V., Gezari, S., et al.\ 2023, \apjl, 955, L6. doi:10.3847/2041-8213/acf216

\bibitem[Zeltyn et al.(2024)]{Zeltyn2024} Zeltyn, G., Trakhtenbrot, B., Eracleous, M., et al.\ 2024, \apj, 966, 85. doi:10.3847/1538-4357/ad2f30

\bibitem[Zhang et al.(2022)]{Zhang2022} Zhang, W.~J., Shu, X.~W., Sheng, Z.~F., et al.\ 2022, \aap, 660, A119. doi:10.1051/0004-6361/202142253

\bibitem[Zhang(2023)]{Zhang2023} Zhang, X.-G.\ 2023, \mnras, 526, 6015. doi:10.1093/mnras/stad3153

\bibitem[Zhou et al.(2019)]{Zhou2019} Zhou, H., Shi, X., Yuan, W., et al.\ 2019, \nat, 573, 83. doi:10.1038/s41586-019-1510-y

\end{thebibliography}
\end{document}